\documentclass[10pt, twocolumn, twoside]{IEEEtran}
\usepackage{graphicx}
\usepackage{color}

\usepackage{amsmath}
\usepackage{amssymb}
\usepackage{arydshln}    
\usepackage{bm}

\usepackage[caption=false,font=footnotesize]{subfig}
\usepackage{cite}

\newtheorem{theorem}{Theorem}

\def\argmin{\mathop{\mathrm{arg\,min}}}

\def\var{\mathop{\mathrm{var}}}
\def\MID{\, | \,} 
\def\P{\mathbb{P}}
\def\Hhat{\widehat{H}}

\def\VMaj{V^{\mbox{\scriptsize \sc maj}}}
\def\VMajN{\VMaj_N}
\def\Vmaj{V^{\mbox{\tiny \sc maj}}}
\def\VmajN{\Vmaj_N}
\def\zetaMaj{\zeta^{\mbox{\scriptsize \sc maj}}}
\def\zetaMajN{\zetaMaj_N}
\def\zetaOr{\zeta^{\mbox{\scriptsize \sc or}}}
\def\zetaOrN{\zetaOr_N}
\def\muOr{\mu^{\mbox{\scriptsize \sc or}}}
\def\muOrN{\muOr_N}
\def\VOr{V^{\mbox{\scriptsize \sc or}}}
\def\VOrN{\VOr_N}
\def\Vor{V^{\mbox{\tiny \sc or}}}
\def\VorN{\Vor_N}
\def\E{\mathbb{E}}
\def\Var{\mathrm{var}}
\newcommand{\LocalI} [1]{{P_{e,#1}^{\rm I}}}
\newcommand{\LocalII}[1]{{P_{e,#1}^{\rm II}}}
\newcommand{\GlobalI} {{P_{E}^{\rm I}}}
\newcommand{\GlobalII}{{P_{E}^{\rm II}}}
\newcommand{\RepP}[2]{a_{#2}^{(#1)}}

\begin{document}

\title{Quantization of Prior Probabilities for \\ Collaborative Distributed Hypothesis Testing%
\thanks{This material is based upon work supported by
  the National Science Foundation under Grants 0729069 and 1101147
  and the Korea Foundation for Advanced Studies (KFAS) Fellowship.
  The material in this paper was presented in part at the
  IEEE Data Compression Conference,
  Snowbird, UT, March 2011.}
}

\author{Joong~Bum~Rhim, Lav~R.~Varshney, and Vivek~K~Goyal%
\thanks{J. B. Rhim (email: jbrhim@mit.edu) and
  V. K. Goyal (email: vgoyal@mit.edu)
  are with the Department of Electrical Engineering and Computer Science and
  the Research Laboratory of Electronics,
  Massachusetts Institute of Technology.}
\thanks{L. R. Varshney (email: varshney@alum.mit.edu) is with the
  IBM Thomas J. Watson Research Center.
}
}

\markboth{Quantization of Prior Probabilities for Collaborative Distributed Hypothesis Testing}
        {Rhim, Varshney, and Goyal}

\maketitle

\begin{abstract}
This paper studies the quantization of prior probabilities, drawn from an ensemble, for distributed detection and data fusion.  Design and performance equivalences between a team of $N$ agents tied by a fixed fusion rule and a more powerful single agent are obtained.  Effects of identical quantization and diverse quantization are compared.  Consideration of perceived common risk enables agents using diverse quantizers to collaborate in hypothesis testing, and it is proven that the minimum mean Bayes risk error is achieved by diverse quantization.  The comparison shows that optimal diverse quantization with $K$ cells per quantizer performs as well as optimal identical quantization with $N(K-1)+1$ cells per quantizer.  Similar results are obtained for maximum Bayes risk error as the distortion criterion.
\end{abstract}


\section{Introduction}
\label{sec:Introduction}

Consider a team of $N$ agents that aims to collaboratively choose between hypotheses $h_0$ and $h_1$ after each agent obtains a noisy observation.
Local decision-making proceeds in parallel, with all agents synchronously observing the object, making hard decisions locally, and sending decisions to the fusion center, without any knowledge of other agents' decisions. The fusion center has some fixed fusion rule known to all agents that generates a global decision based only on local decisions.  Fusion rules that are deterministic and symmetric are of the $L$-out-of-$N$ form whereby the global choice is $h_1$ when $L$ or more agents choose $h_1$.  Examples are the {\sc majority} rule ($L = \lceil \frac{N + 1}{2} \rceil$) and the {\sc or} rule ($L = 1$).
These arise in human affairs, where decision making by juries or committees has been analyzed in economics and political science to understand rational decision rules when the human decision makers possess a common preference for two alternatives~\cite{Austen-SmithB1996}.

With known prior probabilities for the object state,
the local decision rules and the fusion rule can be analyzed and optimized
as standard Bayesian hypothesis testing.
Here, we consider an ensemble of objects, generally uncountably many.
Due to limited memory or limited computational resources, agents are only able to distinguish at most $K$ different categories of objects (and therefore categories of prior probabilities).  As an example, human decision makers are known to think categorically due to the limitation in their information processing capacity~\cite{MacraeBodenhausen2000}.  Due to categorization, the true prior probability of an observed object must be mapped to one of the $K$ discriminable values before each agent performs Bayesian hypothesis testing.  The quality of decisions depends on the individuals' categorization schemes.

In this paper, we study the effect of the categorization and optimization of the categorization.
Since we limit our attention to binary hypothesis testing, a problem from the ensemble is specified by a single scalar $p_0 = \P(H = h_0) = 1 - \P(H = h_1)$.
We model $p_0$ as a realization of a random variable $P_0$, so mapping objects into $K$ categories is equivalent to $K$-level quantization of $P_0$.
Quantization performance is measured by the quality degradation of decisions made based on the quantized prior probability, a Bregman divergence called Bayes risk error~\cite{Varshney2011} averaged over $P_0$, which is called mean Bayes risk error (MBRE). We consider minimum MBRE quantizers in two cases: when all agents use identical quantizers and when they use different quantizers.

A main result of the paper is a large advantage from
diversity among agents. The traditional advantage of having multiple agents in hypothesis testing problems is obtaining more observations about objects: agents observe different noise realizations and reduce the effects of noise by information aggregation.  New here, we show that diversity among agents' mapping schemes for prior probabilities can decrease the chance that inexact prior probabilities lead to wrong decisions. Beneficial diversity helps each agent in cancelling others' wrong decisions so that the quality of a global decision can be improved on average.
Optimal design of the prior-probability quantizers enables $N$ agents with different $K$-level quantizers to perform as well as $N$ agents with identical $(N(K-1)+1)$-level quantizers.

The precise study of quantization of prior probabilities in Bayesian hypothesis testing was recently initiated in~\cite{VarshneyVarshney08}, which focuses on the minimum MBRE quantizer of a single agent. Quantization of prior probabilities to minimize maximum Bayes risk is considered in~\cite{VarshneyVarshney2011}.  The study of quantization of prior probabilities in distributed hypothesis testing by three agents appears in~\cite{RhimVG2011a}, work generalized herein.  We use a single set of Bayes costs as an element of making the agents a \emph{team} in the sense of Marschak and Radner~\cite{MarschakR1972}, i.e., having a common goal.  An alternative is for each agent to have potentially-different Bayes costs. This introduces game-theoretic considerations as described in~\cite{RhimVG2011b}.

Most previous work on the effect of quantization in Bayesian distributed detection is focused on the quantization of observations~\cite{Kassam77,PoorThomas1977,GuptaHero2003} or the communication topology and rates among agents~\cite{KarM2010} and/or to the fusion center~\cite{Tsitsiklis1993,ViswanathanVarshney1997}. We do not consider quantization of observations here, though it may be noted that quantization outside of the system designer's control could be incorporated into the likelihood functions.

The group decision-making model that we are considering is described in Section~\ref{sec:DistributedDetection}. In Section~\ref{sec:Quantization}, we discuss the effect of quantization of prior probabilities on decision making and compare performances of the teams of agents that use identical quantizers and that use different quantizers. Examples of optimal quantizers obtained from our design algorithm are presented in Section~\ref{sec:Example}.  In addition to the mean Bayes risk error, the maximum Bayes risk error is considered in Section~\ref{sec:Minimax}. Section~\ref{sec:Conclusion} concludes the paper.

\section{Distributed Detection and Data Fusion Model}
\label{sec:DistributedDetection}

A binary hypothesis test for a given object is performed by a team of $N$ agents. The object is in state $H = h_0$ with probability $p_0$ and in state $H = h_1$ with probability $1 - p_0$. The agents have a common goal to minimize the cost due to the global decision; the cost for false alarm (misjudgment of $h_0$ as $h_1$) is $c_{10}$ and the cost for missed detection (misjudgment of $h_1$ as $h_0$) is $c_{01}$ for all agents. For simplicity, we consider zero cost for correct decisions.

The observation by Agent $i$, $Y_i$, is governed by the likelihood function $f_{Y_i|H}(y_i \MID H=h_m)$. Each agent makes a hard local decision $\Hhat_i \in \{h_0, h_1\}$ based on its observation and the prior probability it believes. The local decision is transferred to a fusion center to be merged with other agents' decisions by some fixed fusion rule. We consider symmetric fusion rules, which are described as $L$-out-of-$N$ rules for a specific $1 \leq L \leq N$.  This fusion rule returns $\Hhat = h_1$ if at least $L$ agents declare $h_1$; otherwise, it returns $\Hhat = h_0$.

Each individual agent makes the two types of errors with the following probabilities:
\begin{eqnarray*}
    \LocalI{i} & = & \P \left(\Hhat_i = h_1 \MID H = h_0\right), \\
   \LocalII{i} & = & \P \left(\Hhat_i = h_0 \MID H = h_1\right).
\end{eqnarray*}
Then, from the $L$-out-of-$N$ fusion rule, the probabilities of the global decision being in error are:
\begin{eqnarray}
    \GlobalI & = & \sum_{n = L}^{N} \sum_{I \subseteq [N] \atop |I| = n} \prod_{i \in I} \LocalI{i} \prod_{j \in [N] / I} \left(1 - \LocalI{j}\right),
    \label{eq:GlobalI} \\
   \GlobalII & = & \sum_{n = N - L + 1}^{N} \sum_{I \subseteq [N] \atop |I| = n} \prod_{i \in I} \LocalII{i} \prod_{j \in [N] / I} \left(1 - \LocalII{j}\right),
    \label{eq:GlobalII}
\end{eqnarray}
where $[N]$ denotes the set $\{1, 2, \ldots, N\}$.
The Bayes risk is based on the \emph{team} decision:
\begin{align}
    R & = p_0 c_{10} \GlobalI + (1 - p_0) c_{01} \GlobalII.
    \label{eq:BayesRisk}
\end{align}
Agent $a$ minimizes its Bayes risk by adopting the following likelihood ratio test:
\begin{eqnarray}
\label{eq:LikelihoodRatioTest}
\lefteqn{\frac{f_{Y_a | H} (y_a \MID h_1)}{f_{Y_a | H} (y_a \MID h_0)}
\overset{\Hhat_a(y_a) = h_1}{\underset{\Hhat_a(y_a) = h_0}{\gtreqless}}} \\ \notag
& &   \frac{\displaystyle p_0 c_{10} \sum_{I \subseteq [N] / \{a\} \atop |I| = L - 1} \prod_{i \in I} \LocalI{i} \prod_{j \in [N] / I / \{a\}} \left( 1 - \LocalI{j} \right)}{\displaystyle (1 - p_0) c_{01} \sum_{I \subseteq [N] / \{a\} \atop |I| = N - L} \prod_{i \in I} \LocalII{i} \prod_{j \in [N] / I / \{a\}} \left( 1 - \LocalII{j} \right)}.
\end{eqnarray}

Consider the special case of agents making observations through additive Gaussian noise.  That is, being in state $h_m$ sends a signal $s_m \in \mathbb{R}$ to all agents, but Agent $i$ receives the corrupted signal $Y_i = s_m + W_i$, where the noise $W_i$ is assumed to be iid Gaussian with zero mean and variance $\sigma^2$.
Then the likelihood ratio test can be simplified to the decision rule with a decision threshold $\lambda_a$:
\begin{equation*}
    y_a \overset{\Hhat_a(y_a) = h_1}{\underset{\Hhat_a(y_a) = h_0}{\gtreqless}} \lambda_a.
\end{equation*}

By picking the $L$-out-of-$N$ rule as their fusion rule, the agents have symmetry in their decision making under iid additive noise. All agents have the same likelihood ratio test \eqref{eq:LikelihoodRatioTest}, which implies that we can constrain the agents to use identical decision rules.

Limiting attention to identical decision rules considerably simplifies the problem. The solution with an identical local decision rule constraint is asymptotically optimum for the binary hypothesis testing problem~\cite{Tsitsiklis1988}. Furthermore, numerical experience shows that the constraint results in little or no loss of performance for finite $N$ and the optimal fusion rule has the form of an $L$-out-of-$N$ rule~\cite{Tsitsiklis1993}. These results have been obtained in decentralized detection models in which the fusion rule is to be optimized in addition to the local decision rules. Our model has a fixed $L$-out-of-$N$ fusion rule and only local decision rules are subject to be optimized. Numerical experiments show that the restriction to identical decision rules leads to no loss of performance for $N \leq 5$ in the Gaussian likelihoods case (see Sections~\ref{sec:GaussianMajorityExample} and~\ref{sec:OrRule}) and the exponential likelihoods case (see Section~\ref{sec:ExponentialExample}). Thus, we constrain the agents to use identical decision rules in the discussion ahead.

\section{Quantization of Prior Probabilities}
\label{sec:Quantization}
We replace the classical formulation that all agents
know the true prior probability $p_0 = \P(H = h_0)$
with the setting where each Agent $i$ bases its decision rule on $q_i(p_0)$,
where $q_i$ a scalar quantizer on $[0, 1]$.
Such a setting can arise when the team faces an ensemble of binary hypothesis
testing problems and the agents lack the ability to have a different
decision rule for each problem.
The prior probability $p_0$ is thus modeled as a realization of
a random variable $P_0$ with density function $f_{P_0}$.
This section discusses computation of Bayes risk averaged over $P_0$
and the optimization of the quantizers applied to the prior probability.

In order to maximize team performance, all agents must coordinate by
sharing their quantized prior probabilities.  The following steps describe the collaborative
decision making process.
\begin{enumerate}
    \item \textit{Quantizer design:}  The $K$-level quantizers $q_1, \ldots, q_N$ are designed.  These quantizers remain fixed for all time.
    \item \textit{Coordination:}  Agents encounter an object whose prior probability is $p_0$.  Each agent applies its quantizer to the prior probability and sends the output $q_i(p_0)$ to all other agents as a part of collaboration.
    \item \textit{Decision rule design:}  Agents design the best common decision rule based on $q_1(p_0), \ldots, q_N(p_0)$.
    \item \textit{Signal observation and decision making:}  Each agent observes a noisy signal $Y_i$ and applies the decision rule to make a local decision $\Hhat_i$.
    \item \textit{Decision fusion:}  All local decisions $\Hhat_i$ are fused to produce a global decision $\Hhat$.
\end{enumerate}
Quantizer design and decision rule design are detailed in this section.  The operations of local decision making and global decision fusion have already been discussed.  Coordination simply involves communication.

The Bayes risk incurred when each agent bases
its local decision on its quantized prior probability is denoted $R_M(q_1(p_0), \ldots, q_N(p_0))$.\footnote{Note that by
coordination, each agent knows other agents' quantized prior probabilities.}
This is called \emph{mismatched Bayes risk} to contrast to the true Bayes risk
$R(p_0)$ if the agents know the true prior probability $p_0$. The Bayes risk error is defined as
\begin{equation}
    d(p_0, q_1(p_0), \ldots, q_N(p_0)) = R_M(q_1(p_0), \ldots, q_N(p_0)) - R(p_0). \nonumber
\end{equation}
We compute the mean Bayes risk error (MBRE) as the distortion of the quantizers for prior probabilities, which measures the average performance over all $p_0$:
\begin{align}
    D & = \E[d(P_0, q_1(P_0), \ldots, q_N(p_0)] \nonumber \\
    & = \int_{0}^{1} d(p_0, q_1(p_0), \ldots, q_N(p_0)) f_{P_0}(p_0) \, dp_0.
    \nonumber
\end{align}
Minimum MBRE quantizers are to be designed within this model.

\subsection{Single Agent}    \label{sec:SingleAgentQuantization}

Let us first review the case of $N = 1$~\cite{VarshneyVarshney08}.
The optimal choice of $K$ values for quantized prior probabilities is described by the minimum MBRE quantizer
\begin{equation*}
    q^{*}_1 = \argmin_{q_1} \E[d(P_0, q_1(P_0))],
\end{equation*}
where the distortion measure of the single quantizer $q_1$ is the Bayes risk error
\begin{eqnarray*}
    \lefteqn{d(p_0, q_1(p_0))  = R_M(q_1(p_0)) - R(p_0)} \\
    & = & p_0 c_{10} p_0 \GlobalI(q_1(p_0)) + (1 - p_0) c_{01} \GlobalII(q_1(p_0)) \\
    &   & \quad -\, p_0 c_{10} \GlobalI(p_0) - (1 - p_0) c_{01} \GlobalII(p_0).
\end{eqnarray*}

\begin{theorem}\textup{\cite[Thm.~1]{VarshneyVarshney08}}
    The Bayes risk error $d(p_0, a)$ is nonnegative and only equal to zero when $p_0 = a$. As a function of $p_0 \in (0, 1)$, it is continuous and strictly convex for all $a$.
    \label{thm:Varshney1}
\end{theorem}

\begin{theorem}\textup{\cite[Thm.~2]{VarshneyVarshney08}}
    For any deterministic likelihood ratio test $\Hhat(\cdot)$, as a function of $a \in (0, 1)$ for all $p_0$, the Bayes risk error $d(p_0, a)$ has exactly one stationary point, which is a minimum.
    \label{thm:Varshney2}
\end{theorem}

Due to the strict convexity of $d(p_0, a)$ in $p_0$ for all $a$, quantizers that satisfy necessary conditions for MBRE optimality are regular; quantization cells are subintervals $\mathcal{R}_1 = [0, b_1), \mathcal{R}_2 = [b_1, b_2), \ldots, \mathcal{R}_K = [b_{k - 1}, 1]$, and each representation point $a_k$ is in $\mathcal{R}_k$.  The necessary conditions for the optimality of a quantizer for $f_{P_0}(p_0)$ are now described.

A nearest neighbor condition describes an expression for the cell boundaries $\{b_k\}$ for fixed representation points $\{a_k\}$.  Between two consecutive representation points $a_k$ and $a_{k + 1}$, the cell boundary $b_k$ needs to separate $x_1$ such that $d(x_1, a_k) < d(x_1, a_{k + 1})$ and $x_2$ such that $d(x_2, a_k) > d(x_2, a_{k + 1})$. Thus, $b_k = b^{*}$ is obtained from the condition
\begin{equation}
\label{eq:NearestNeighbor}
    d(b^{*}, a_{k}) = d(b^{*}, a_{k + 1}).
\end{equation}
The point is found to be
\begin{equation*}
    b^{*} = \frac{c_{01}\left( \GlobalII(a_{k + 1}) - \GlobalII(a_{k}) \right)}{c_{01} \left( \GlobalII(a_{k + 1}) - \GlobalII(a_k)\right) - c_{10} \left( \GlobalI(a_{k + 1}) - \GlobalI(a_k) \right)}.
\end{equation*}

A centroid condition describes optimal representation points for fixed quantization cells. The MBRE is expressed as the sum of integrals over quantization regions
\begin{equation}
    D = \sum_{k = 1}^{K} \int_{\mathcal{R}_k} d(p_0, a_k) f_{P_0}(p_0) \, dp_0, \nonumber
\end{equation}
and the minimization may be performed for each cell separately. The representation point $a_k = a^{*}$ of the cell $\mathcal{R}_k$ is chosen from the optimization problem
\begin{equation*}
    a^{*} = \argmin_{a} \left\{ \int_{\mathcal{R}_k} d(p_0, a) f_{P_0}(p_0) \, dp_0 \right\}.
\end{equation*}
Since Bayes risk error is a Bregman divergence~\cite{Varshney2011}, the unique minimizer is the centroid of the region $\mathcal{R}_k$:
\begin{equation*}
    a^{*} = \frac{\int_{\mathcal{R}_k} p_0 f_{P_0}(p_0) \, dp_0}{\int_{\mathcal{R}_k} f_{P_0}(p_0) \, dp_0}.
\end{equation*}

The minimum MBRE quantizer $q^{*}_1$ can be found by the iterative Lloyd--Max algorithm, which alternates between the nearest neighbor and the centroid conditions.

\subsection{Identical Quantizers}    \label{sec:IdenticalQuantization}

Consider, for $N > 1$, that all agents use the same quantizer $q$ for prior probabilities; coordination informs all agents of the identicality. For an object with prior probability $p_0$, all agents use $p' = q(p_0)$ as the object's prior. The agents will incur the mismatched Bayes risk
\begin{equation}
    R_{M} = p_0 c_{10} \GlobalI + (1 - p_0) c_{01} \GlobalII
    \label{eq:MismatchedBayesRisk}
\end{equation}
due to their decisions, but what they minimize is \textit{perceived Bayes risk}:
\begin{equation}
    R_{P} = p' c_{10} \GlobalI + (1 - p') c_{01} \GlobalII.
    \label{eq:PerceivedBayesRisk}
\end{equation}
The probabilities $\GlobalI$ and $\GlobalII$ in (\ref{eq:MismatchedBayesRisk}) and (\ref{eq:PerceivedBayesRisk}) are identical. In other words, $\GlobalI$ and $\GlobalII$ are determined from the decision rules that minimize the perceived Bayes risk (\ref{eq:PerceivedBayesRisk}) and applied to compute the mismatched Bayes risk (\ref{eq:MismatchedBayesRisk}).

A general way to find local decision rules is to directly optimize \eqref{eq:GlobalI} and \eqref{eq:GlobalII}, but these are generally complicated functions.  One way to simplify the analysis for additive observation noise models is to find a decision-making model with a single agent whose performance is the same as the team of the multiple agents.  By the same performance, we mean that the single agent uses the same optimal decision rule and its Type I and Type II error probabilities are respectively equal to $\GlobalI$ and $\GlobalII$ when it uses the same quantizer $q$ for prior probabilities.

\begin{theorem}
    Assume that the unquantized prior probability is known. Consider $N$ agents that perform group decision-making with observations corrupted by additive noises $W_1, \ldots, W_N$. For convenience, index the agents in descending order of the realizations of the noises: $W_{(1)} \geq W_{(2)} \geq \cdots \geq W_{(N)}$. When their decisions are fused by the $L$-out-of-$N$ rule, their performance is the same as that of a single agent having the same Bayes costs if its observation is corrupted by the $L$th largest additive noise $V = W_{(L)}$.
    \label{thm:EquivalentModel}
\end{theorem}
\begin{IEEEproof}
    The global decision is the same as the decision of Agent $L$ because all agents adopt the same decision threshold $\lambda$. If Agent $L$ declares $h_0$, then Agents $L + 1, \ldots, N$, whose observations are smaller than or the same as that of Agent $L$, also declare $h_0$. Since at least $N - L + 1$ agents send $h_0$, the fusion rule gives $h_0$ as the global decision. If Agent $L$ declares $h_1$, then Agents $1, \ldots, L - 1$, whose observations are larger than or at least the same as that of Agent $L$, also declare $h_1$ and the global decision is $h_1$.

    As such, the Bayes risk (\ref{eq:BayesRisk}) can be rewritten as
    \begin{eqnarray}
        R & = & p_0 c_{10} \GlobalI + (1 - p_0) c_{01} \GlobalII \nonumber \\
        & = & p_0 c_{10} \P (W_{(L)} + s_0 \geq \lambda \MID H = h_0) \nonumber \\
        &   & \quad + (1 - p_0) c_{01} \P (W_{(L)} + s_1 < \lambda \MID H = h_1).
        \label{eq:BayesRisk2}
    \end{eqnarray}
    If we consider a new single-agent problem with additive noise $V = W_{(L)}$, then the Bayes risk of the single agent is equal to (\ref{eq:BayesRisk2}). Therefore, the optimal decision rule of the single agent is equal to that of the multiple agents, and the single agent obtains the same performance as the team of multiple agents.
\end{IEEEproof}

When the $W_i$ are iid continuous random variables with pdf $f_{W}$ and cdf $F_{W}$, the random variable $V = W_{(L)}$ is well understood from the theory of order statistics~\cite{DavidNagaraja2003}. The pdf of $V$ is
\begin{align*}
&f_{V}(v) = \\ \notag
&\quad \frac{N!}{(N - L)! (L - 1)!} F_{W}^{N - L}(v) \left[1 - F_W(v)\right]^{L - 1} f_W(v).
\end{align*}
Thus, we only need to consider a single agent with a different additive noise $V$, no matter how many agents there are. The Bayesian decision rule of the single agent that observes $Y = s_m + V$ is given by
\begin{equation*}
    \frac{f_{Y | H} (y \MID h_1)}{f_{Y | H} (y \MID h_0)} = \frac{f_V(y - s_1)}{f_V(y - s_0)} \overset{\Hhat(y) = h_1}{\underset{\Hhat(y) = h_0}{\gtreqless}}  \frac{p_0 c_{10}}{(1 - p_0) c_{01}},
\end{equation*}
which can also be used by the multiple agents.

\begin{theorem}
    If multiple agents use identical quantizers for prior probabilities, then their optimal quantizer is equal to the optimal quantizer in the equivalent single-agent model.
    \label{thm:IdenticalQuantizer}
\end{theorem}
\begin{IEEEproof}
    Since Theorem~\ref{thm:EquivalentModel} is valid for any prior probability, Theorem~\ref{thm:EquivalentModel} holds for quantized prior probabilities whenever the multiple agents and the single agent quantize the prior probability to the same value. Thus, their average performances are also the same if they use the same quantizer $q$ for prior probability. As a result, the single agent can achieve the minimum MBRE by adopting the optimal quantizer of the multiple agents and vice versa.
\end{IEEEproof}

The analysis and design of the minimum MBRE quantizer in a single-agent model in~\cite{VarshneyVarshney08} can be applied to the multiple-agent model without any change except the noise model.

\subsection{Diverse Quantizers}    \label{sec:DiverseQuantization}

Now consider the setting where each agent may have its own quantizer for prior probabilities. For an object with prior probability $p_0$, Agent $i$ believes that the probability of the object being in state $h_0$ is $p^{(i)} = q_i(p_0)$, which may be different from Agent $j$'s quantization $p^{(j)} = q_j(p_0)$.  This diversity may allow the agents to improve the quality of their decisions as a team.

In order to take advantage of diversity, the agents need to collaborate in their decision making.  The coordination phase of the collaborative decision making process is key.  When agents use identical quantizers, all agents consider the same perceived Bayes risk \eqref{eq:PerceivedBayesRisk} even if they do not communicate, but here collaboration must be established.

Agents may try to minimize their own perceived Bayes risks so as to give the best decisions for all agents.  When all agents quantize the prior probability to different values, however, they have different perceived Bayes risks:
\[
R_{\rm P}^{(i)} = p^{(i)} c_{10} \GlobalI + \left(1 - p^{(i)}\right) c_{01} \GlobalII \neq R_{\rm P}^{(j)}.
\]
If all agents individually optimize their decision rules, the resulting performance is not as good as their best~\cite{RhimVG2011b}.

All agents will incur the same cost as a result of their decisions: $c_{10}$ if their team misreads $h_0$ as $h_1$ and $c_{01}$ if their team misreads $h_1$ as $h_0$. This motivates them to collaborate by sharing a common goal that replaces their individual perceived Bayes risks.  As the common goal, we introduce a weighted sum of the perceived Bayes risks called \textit{perceived common risk}:
\begin{equation*}
    R_{\rm PC} = \sum_{i = 1}^{N} u_i R_{\rm P}^{(i)},
\end{equation*}
where $u_i$ are constants that satisfy $\sum_{i = 1}^{N} u_i = 1$ and $u_i > 0$ for all $i \in [N]$.

The introduction of the perceived common risk allows us to treat identical-quantizer and diverse-quantizer settings on a common footing.  The minimum perceived-common-risk decision rule can be used since, if all agents use identical quantizers like in Section~\ref{sec:IdenticalQuantization}, their perceived common risk is equal to their perceived Bayes risks:
\begin{eqnarray*}
    R_{\rm PC} & = & \sum_{i = 1}^{N} u_i \left( p^{(i)} c_{10} \GlobalI + \left(1 - p^{(i)}\right) c_{01} \GlobalII \right) \\
    & = & \sum_{i = 1}^{N} u_i \left( p' c_{10} \GlobalI + \left(1 - p'\right) c_{01} \GlobalII \right) \\
    & = & p' c_{10} \GlobalI + \left(1 - p'\right) c_{01} \GlobalII
    \ = \ R_{\rm P}.
\end{eqnarray*}

It may seem unintuitive to use the perceived common risk as a criterion of decision making because decision rules, which are determined based on the perceived common risk, may not be optimal for some prior probability $p_0$. However, it is the Bayes risk averaged over $P_0$ that measures the performance of the team. Thus, a set of quantizers that leads to good decisions on average is essential for the agents to use the perceived common risk as a common goal of minimization.

Design of such quantizers has high computational complexity. In Section~\ref{sec:IdenticalQuantization}, a single quantizer is easily designed by the iterative Lloyd--Max algorithm, which utilizes independence among endpoints of quantization cells in the nearest neighbor step (optimizing cell boundaries) and independence among representation points in the centroid step (optimizing representation points). On the other hand, multiple quantizers that define different quantization cells do not have such independencies.

\begin{figure}
    \centering
    \includegraphics[width=3in]{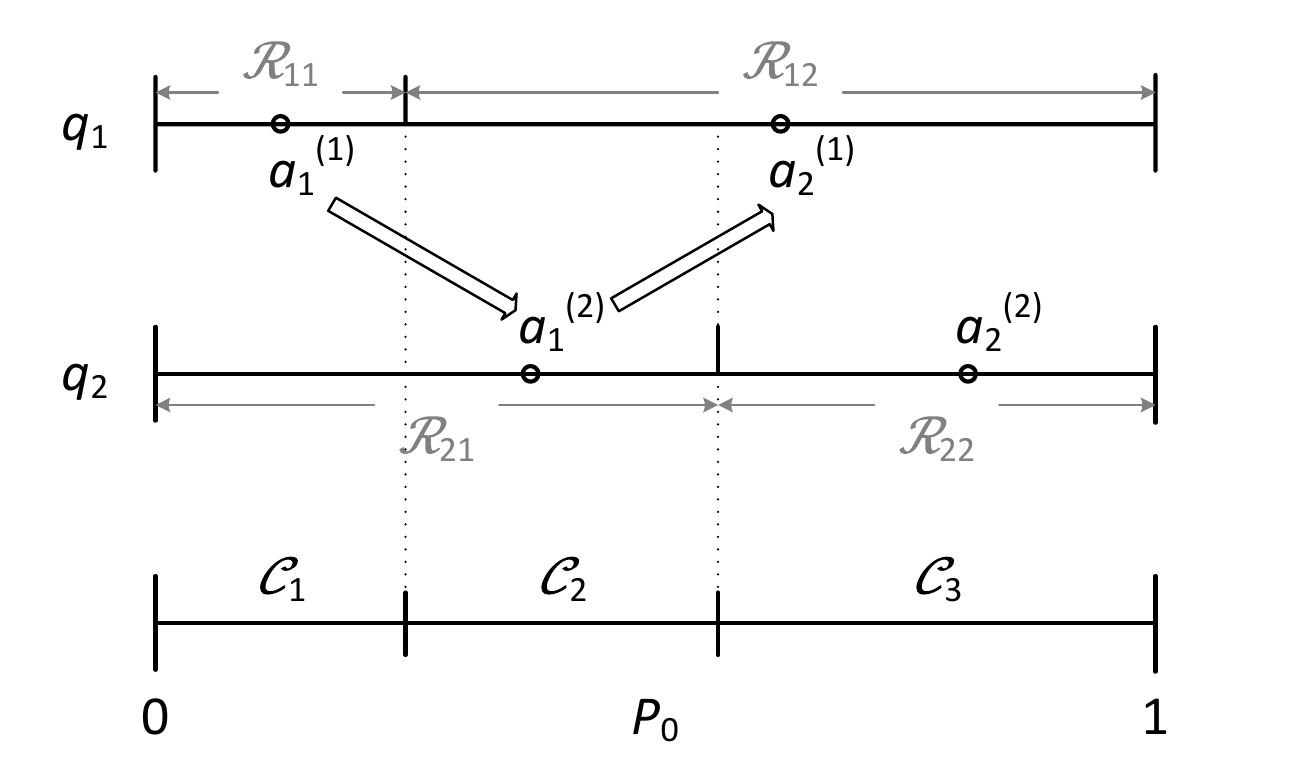}
    \caption{An example that shows how the dependency among representation points propagates to break the independence between representation points of two different cells $\mathcal{R}_{11}$ and $\mathcal{R}_{12}$.}
    \label{fig:QuantizerDependency}
\end{figure}

For the simple example of Fig.~\ref{fig:QuantizerDependency}, the cell $\mathcal{R}_{21}$ of the quantizer $q_2$ affects the decisions for objects whose prior probabilities are within the interval $\mathcal{C}_1$ or $\mathcal{C}_2$. The team performance in the interval $\mathcal{C}_1$ is affected by the representation points $\RepP{1}{1}$ and $\RepP{2}{1}$ and by $\RepP{1}{2}$ and $\RepP{2}{1}$ in the interval $\mathcal{C}_2$. Hence, we can observe that dependency is propagated through the representation points: a choice of $\RepP{1}{2}$ depends on a choice of $\RepP{2}{1}$, which depends on a choice of $\RepP{1}{1}$. In order to avoid such complexity, we introduce an indirect method to optimize diverse quantizers.

Consider a team of $N$ agents that respectively use different $K$-level quantizers $q_1, \ldots, q_N$ and another team of $N$ agents that all use an identical $K_S$-level quantizer $q_S$.

\begin{theorem}
    \label{thm:EquivalentQuantizer}
    A set of $N$ different $K$-level quantizers $\{q_1, \ldots, q_N\}$ and a $K_S$-level quantizer $q_S$ result in the same perceived common risk if, for all $p_0 \in [0, 1]$,
    \begin{equation}
        \sum_{i = 1}^{N} u_i q_i(p_0) = q_S(p_0).
        \label{eq:EquivalentQuantizerCondition}
    \end{equation}
\end{theorem}
\begin{IEEEproof}
    For an object with prior probability $p_0$, agents respectively using $\{q_1, \ldots, q_N\}$ make decisions based on individually quantized prior probabilities $p^{(i)} = q_i(p_0)$. Their perceived common risk is:
    \begin{align}
        R_{\rm PC,1} & = \sum_{i = 1}^{N} u_i \left( p^{(i)} c_{10} \GlobalI + \left(1 - p^{(i)}\right) c_{01} \GlobalII \right) \nonumber \\
        & = \left( \sum_{i = 1}^{N} u_i p^{(i)} \right) c_{10} \GlobalI + \left( \sum_{i = 1}^{N} u_i \left(1 - p^{(i)}\right) \right) c_{01} \GlobalII.
        \label{eq:PerceivedCommonRisk1}
    \end{align}
    For agents using the common quantizer $q_S$, with quantized prior probability $p' = q_S(p_0)$, their perceived common risk is
    \begin{equation}
        R_{\rm PC,2} = p' c_{10} \GlobalI + (1 - p') c_{01} \GlobalII.
        \label{eq:PerceivedCommonRisk2}
    \end{equation}

    The perceived common risks (\ref{eq:PerceivedCommonRisk1}) and (\ref{eq:PerceivedCommonRisk2}) only depend on the decision rules used by $N$ agents.  If there exists a constant $t$ that satisfies
\[R_{\rm PC,1} = t \cdot R_{\rm PC,2}\]
for any $N$-tuple of decision rules, then the $N$-tuple that minimizes $R_{\rm PC,1}$ also minimizes $R_{\rm PC,2}$. If the proportionality property holds for all $p_0 \in [0, 1]$, then the agents in both cases always use the same decision rules. Therefore, the set of different quantizers $\{q_1, \ldots, q_N\}$ and the set of identical quantizers $\{q_S, \ldots, q_S\}$ will cause the same Bayes risk on any problem in the ensemble.

    For any $N$-tuple and any $p_0$, $R_{\rm PC,1} = t \cdot R_{\rm PC,2}$ if and only if $\sum_{i = 1}^{N} u_i p^{(i)} = t p'$ and $\sum_{i = 1}^{N} u_i \left(1 - p^{(i)}\right) = t (1 - p')$. From the fact that $\sum_{i = 1}^{N} u_i = 1$, the constant $t$ can only be 1, and the condition is simplified to
    \begin{equation*}
        \sum_{i = 1}^{N} u_i q_i(p_0) = q_S(p_0), \qquad \mbox{for all $p_0 \in [0, 1]$}.
    \end{equation*}
    This simple condition comes from the fact that the perceived common risk is a weighted sum of the perceived Bayes risks.
\end{IEEEproof}

\begin{theorem}
    A team of $N$ agents individually using $N$ diverse $K$-level quantizers can achieve the minimum mean Bayes risk error that they can achieve when they use the same $(N (K - 1) + 1)$-level quantizer.
    \label{thm:GuaranteedLevel}
\end{theorem}
\begin{IEEEproof}
    When the agents use an identical $K_S$-level quantizer, the optimal quantizer $q_S$ is always a regular quantizer~\cite{VarshneyVarshney08}; each quantization cell $\mathcal{C}_k$ is an interval and its representation point $x_k$ is within the interval. For each cell $\mathcal{C}_k$ of $q_S$, $k = 1, \ldots, K_S$, the condition (\ref{eq:EquivalentQuantizerCondition}) is expressed by an equation
    \begin{equation}
        x_k = \sum_{i = 1}^{N} a^{(i)} u_i, \nonumber
    \end{equation}
    where $a^{(i)}$ is one of the representation points of $K$-level quantizer $q_i$, $i = 1, \ldots, N$.  Overall, we have a total of $K_S$ equations, which are described in the following matrix form:
    \begin{equation}
        \left[ \begin{array}{c} x_1 \\ \vdots \\ x_{K_S} \end{array} \right]
        = \mathbf{A} \left[ \begin{array}{c} u_1 \\ \vdots \\ u_N \end{array} \right],
        \label{eq:EquivalentQuantizerConditionMatrix}
    \end{equation}
    where $\mathbf{A}$ is a $K_S \times N$ matrix that has $\left[ \RepP{1}{k_{j1}} \  \RepP{2}{k_{j2}} \  \cdots \  \RepP{N}{k_{jN}} \right]$ as its $j$th row and $\RepP{i}{k_{ji}}$ is the representation point of the $k_{ji}$th cell of $q_i$ for any $1 \leq k_{ji} \leq K$.

    There exists an $\mathbf{A}$ that satisfies (\ref{eq:EquivalentQuantizerConditionMatrix}) if and only if there exists a set of $N$ different $K$-level quantizers $q_1, \ldots, q_N$ that are equivalent to $q_S$. In (\ref{eq:EquivalentQuantizerConditionMatrix}), the vector $[u_1 \ u_2 \ \cdots \ u_N]^T$ is a given parameter and the vector $[x_1 \ x_2 \ \cdots \ x_{K_S}]^T$ is uniquely determined by the optimization of $q_S$; the matrix $\mathbf{A}$ consists of unknown parameters (i.e., representation points of $q_i$) to be determined. However, if $\mathbf{A}$ has any linearly dependent row, the representation points may not exist. On the other hand, if $\mathbf{A}$ has only linearly independent rows and $K_S \leq N K$, then we can solve (\ref{eq:EquivalentQuantizerConditionMatrix}) to find all the representation points of $q_i$.

    The maximum number of linearly independent rows of $\mathbf{A}$ is $N(K - 1) + 1$:
    \begin{equation}
        \mathbf{A} = \left[
            \begin{array}{cccc}
                \RepP{1}{1} & \RepP{2}{1} & \cdots & \RepP{N}{1} \\
                \hdashline[1pt/3pt]
                \RepP{1}{2} & \RepP{2}{1} & \cdots & \RepP{N}{1} \\
                \RepP{1}{1} & \RepP{2}{2} & \cdots & \RepP{N}{1} \\
                \vdots & \vdots & \ddots & \vdots \\
                \RepP{1}{1} & \RepP{2}{1} & \cdots & \RepP{N}{2} \\
                \hdashline[1pt/3pt]
                \RepP{1}{3} & \RepP{2}{1} & \cdots & \RepP{N}{1} \\
                \RepP{1}{1} & \RepP{2}{3} & \cdots & \RepP{N}{1} \\
                \vdots & \vdots & \ddots & \vdots \\
                \RepP{1}{1} & \RepP{2}{1} & \cdots & \RepP{N}{3} \\
                \hdashline[1pt/3pt]
                \vdots & \vdots & \ddots & \vdots \\
                \hdashline[1pt/3pt]
                \RepP{1}{N} & \RepP{2}{1} & \cdots & \RepP{N}{1} \\
                \RepP{1}{1} & \RepP{2}{N} & \cdots & \RepP{N}{1} \\
                \vdots & \vdots & \ddots & \vdots \\
                \RepP{1}{1} & \RepP{2}{1} & \cdots & \RepP{N}{N} \\
            \end{array}
        \right]. \nonumber
    \end{equation}
Any other row $\left[ \RepP{1}{k_{1}} \  \RepP{2}{k_{2}} \  \cdots \  \RepP{N}{k_{N}} \right]$ is a linear combination of the $N(K - 1) + 1$ rows. Thus, the existence of $\{q_1, \ldots, q_N\}$ that is equivalent to $q_S$ is only guaranteed when $K \leq K_S \leq N(K - 1) + 1$.

A quantizer with greater $K_S$ does not increase mean Bayes risk error~\cite{VarshneyVarshney08}. Therefore, the minimum mean Bayes risk error of agents using diverse $K$-level quantizers is upper bounded by that of agents using the identical $(N (K - 1) + 1)$-level quantizer.
\end{IEEEproof}

The optimal set of $K$-level quantizers for $N$ agents is always a set of diverse quantizers if $K > 1$. The optimal set of quantizers can be designed by a two-step algorithm. The first step is to design the optimal $(N (K - 1) + 1)$-level quantizer $q_S$ to be commonly used by $N$ agents, e.g.\ with the Lloyd--Max algorithm~\cite{VarshneyVarshney08}.

The second step is to disassemble the quantizer $q_S$ into $N$ different $K$-level quantizers $q_1, \ldots, q_N$ that lead to the same mean Bayes risk. For convenience, we consider the $K$-level quantizers whose quantization cells are intervals, i.e., the cell $\mathcal{R}_{ik}$ is defined to be $[b_{k-1}, b_k)$ for any Agent $i$ and $1 \leq k \leq K$.

In order to satisfy (\ref{eq:EquivalentQuantizerCondition}), each cell boundary of $q_S$ needs to be a cell boundary of at least one of $q_1, \ldots, q_N$ because, for any cell boundary $y$ of $q_S$ and for any $x_1 < y$ and $x_2 > y$, $\sum_{i = 1}^{N} u_i q_i(x_1) = q_S(x_1) \neq q_S(x_2) = \sum_{i = 1}^{N} u_i q_i(x_2)$. Also, any two or more quantizers out of $q_1, \ldots, q_N$ cannot have a common cell boundary since $q_S$ has $N (K - 1) + 1$ cells. As a result, each cell boundary of $q_S$, except $0$ and $1$, needs to be a cell boundary of exactly one of $q_1, \ldots, q_N$.

Therefore, we determine cell boundaries of $q_1, \ldots, q_N$ by splitting the cell boundaries of $q_S$ into $N$ sets by the following conditions:
\begin{align}
    \bigcup_{i = 1}^{N} B_i & = B_S, \nonumber \\
    B_i \cap B_j & = \{0, 1\}, \qquad \mbox{for all $i$ and all $j \neq i$},
    \label{eq:BoundaryCondition}
\end{align}
where $B_S$ is the set of cell boundaries of $q_S$ and $B_i$ are the sets of cell boundaries of $q_i$; $|B_S| = N (K - 1) + 2$ and $|B_i| = K + 1$. The sets $B_1, \ldots, B_N$ that satisfy (\ref{eq:BoundaryCondition}) are not unique.

Representation points of the cells of $q_1, \ldots, q_N$ are determined after cell boundaries of the quantizers are fixed. We have $K_S = N (K - 1) + 1$ equations of representation points of $q_1, \ldots, q_N$ that describe the condition (\ref{eq:EquivalentQuantizerCondition}). Once we find the representation points that satisfy all the equations, then the quantizers are minimum MBRE diverse quantizers for the $N$ agents. The optimal identical quantizer $q_S$ is always regular because the optimal quantizer for single agent is regular, but the optimal diverse quantizers $q_1, \ldots, q_N$ need not be regular.

The total number of the representation points is $N K$, but we have $N - 1$ less equations than what we need to uniquely determine the representation points. Furthermore, a different arrangement of cell boundaries changes the equations and, consequently, proper representation points. Therefore, optimal diverse $K$-level quantizers are not unique; any choice leads to the same result because Theorem~\ref{thm:EquivalentQuantizer} shows that any set of the optimal quantizers causes the same mean Bayes risk error as $q_S$ does.

\subsection{Comparison to Team-Oblivious Agents} \label{sec:Comparison}
We have discussed teams of agents that are aware of the existence of the other $N - 1$ agents and the $L$-out-of-$N$ fusion rule.  Let us now consider the case when the agents do not know $N$ and $L$; individual agents maximize their own probability of being correct.  Agent $i$ considers the Bayes risk
\begin{equation*}
    R = p_0 c_{10} \LocalI{i} + (1 - p_0) c_{01} \LocalII{i}.
\end{equation*}
All agents choose their quantizers as if they are single agents with additive noise $W$ drawn from $f_W$.  Consequently, they have identical quantizers $q$ because they have the same Bayes costs $c_{10}$ and $c_{01}$.

Their quantizers are obviously not optimal.  In Section~\ref{sec:IdenticalQuantization}, it has been shown that their optimal identical quantizers need to be designed based on the additive noise $V = W_{(L)}$, whose density function is different from $f_W$, in a single-agent model.  Furthermore, any identical quantizers can be improved by transformation of diverse quantizers, as in Section~\ref{sec:DiverseQuantization}.  Therefore, for each agent, minimization of the MBRE of its own decision is not the best strategy as an optimal team member and its optimal behavior depends on $N$ and $L$.  The effect of $N$ and $L$ on the design of optimal quantizers is reflected by the equivalent single-agent model in Theorem~\ref{thm:EquivalentModel}.

\section{Examples}    \label{sec:Example}

Throughout this section, assume that the prior probability $p_0$ of an object being in state $h_0$ is drawn from the uniform distribution $f_{P_0}(p_0) = 1$ for all $p_0 \in [0, 1]$.

\subsection{Gaussian Likelihoods, {\sc majority} Rule}    \label{sec:GaussianMajorityExample}

Suppose the object sends the signal $s_0 = 0$ in state $h_0$ and $s_1 = 1$ in state $h_1$, and each agent receives the signal corrupted by additive iid noise $W_i$ with cdf $F_W$ and pdf $f_W$ where
\begin{equation}
    f_W(w) = \frac{1}{\sqrt{2 \pi \sigma^2}} e^{-w^2 / 2 \sigma^2}. \nonumber
\end{equation}
Each likelihood $f_{Y_i|H}$ is thus Gaussian with mean $H$.
The {\sc majority} rule is used to fuse local decisions made by $N$ agents. We assume that $N$ is an odd number so that voting never results in a tie.

From Theorem~\ref{thm:EquivalentModel}, the equivalent single-agent model has an additive noise $\VMajN$ with pdf given by
\begin{equation*}
f_{\VmajN}(v) = 
\frac{N!}{\left(\frac{N - 1}{2}! \right)^2} F_{W}^{(N - 1) / 2}(v) \left[1 - F_W(v)\right]^{(N - 1) / 2} f_W(v).
\end{equation*}
The mean of $\VMajN$ is zero and the variance of $\VMajN$ is proportional to that of $W_i$:
\begin{align}
    & \Var(\VMajN) \nonumber \\
    & = \int_{- \infty}^{\infty} v^2 \frac{N!}{\left( \frac{N - 1}{2}! \right)^2} F_W^{\frac{N - 1}{2}}(v) \left[1 - F_W(v)\right]^{\frac{N - 1}{2}} f_W(v) \, dv \nonumber \\
    & = \int_{- \infty}^{\infty} v^2 \frac{N!}{\left( \frac{N - 1}{2}! \right)^2} F_{\mathcal{N}}^{\frac{N - 1}{2}}\left(\textstyle \frac{v}{\sigma}\right) \left[1 - F_{\mathcal{N}}\left(\textstyle \frac{v}{\sigma}\right)\right]^{\frac{N - 1}{2}} f_W(v) \, dv \nonumber \\
    & = \sigma^2 \int_{- \infty}^{\infty} z^2 \frac{N!}{\left( \frac{N - 1}{2}! \right)^2} F_{\mathcal{N}}^{\frac{N - 1}{2}}(z) \left[1 - F_{\mathcal{N}}(z)\right]^{\frac{N - 1}{2}} f_{\mathcal{N}}(z) \, dz \nonumber \\
    & \triangleq \sigma^2 \zetaMajN,
    \label{eq:MajorityRuleVariance}
\end{align}
where $f_{\mathcal{N}}$ and $F_{\mathcal{N}}$ denote the pdf and cdf of
a standard normal random variable,
and $z = v / \sigma$. The factor $\zetaMajN$ is the variance of the median of $N$ iid standard normal random variables.
The means and variances of Gaussian order statistics have been studied extensively~\cite{Tippett1925, Teichroew1956, BoseGupta1959}, and the value of $\zetaMajN$ for any $1 \leq N \leq 20$ can be found in~\cite{Teichroew1956}.
For example, $\zetaMaj_3 = 0.4487$ and $\zetaMaj_5 = 0.2863$.

This equivalent single-agent model is considered for optimization of identical $(N(K - 1) + 1)$-level quantizers.  An optimal set of diverse $K$-level quantizers for the agents can be designed from the optimal $(N(K-1) + 1)$-level quantizer.

\begin{figure}
    \centering{
    \subfloat[$K=1$]{\includegraphics[width=1.8in]{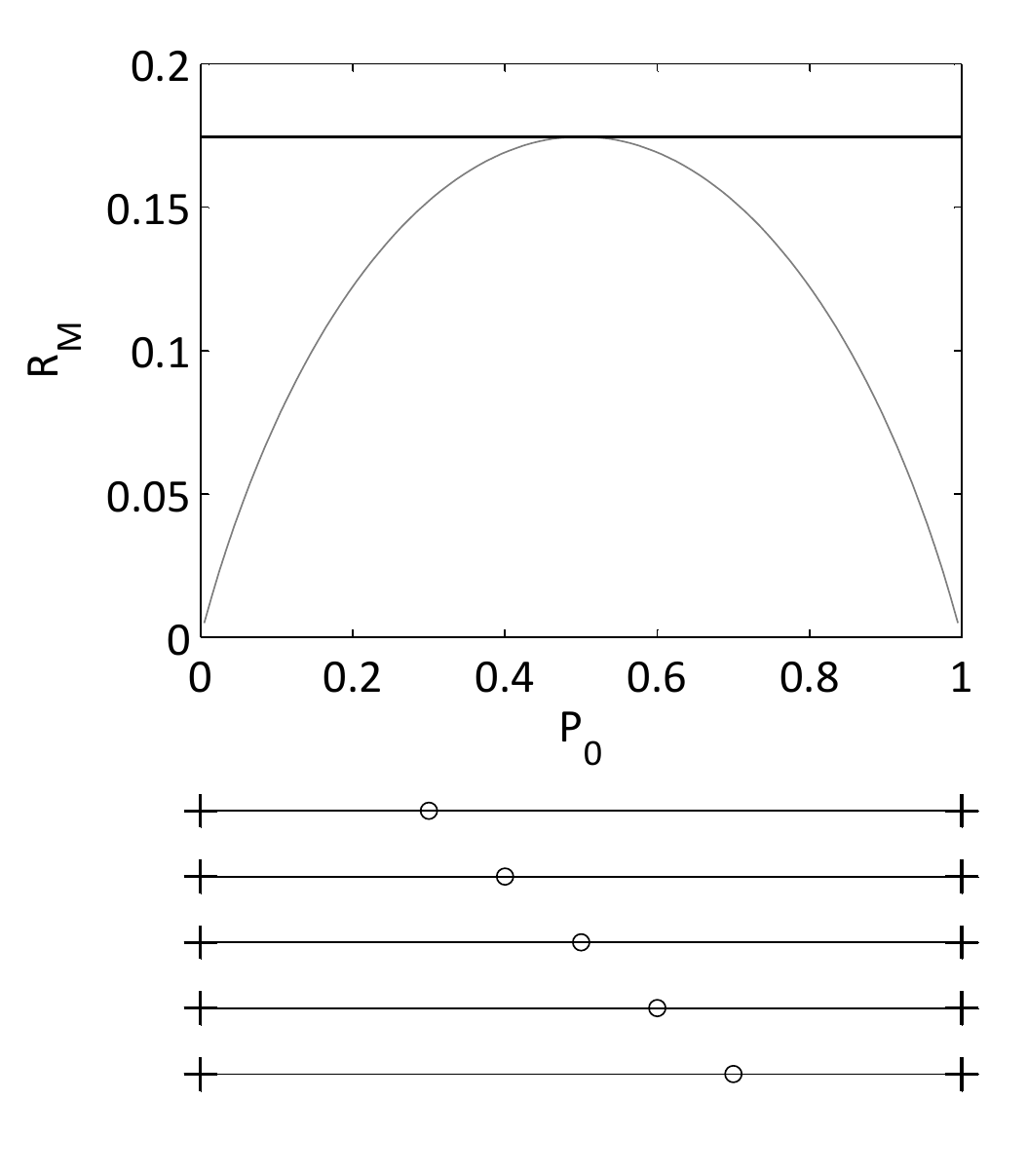}}
    \subfloat[$K=2$]{\includegraphics[width=1.8in]{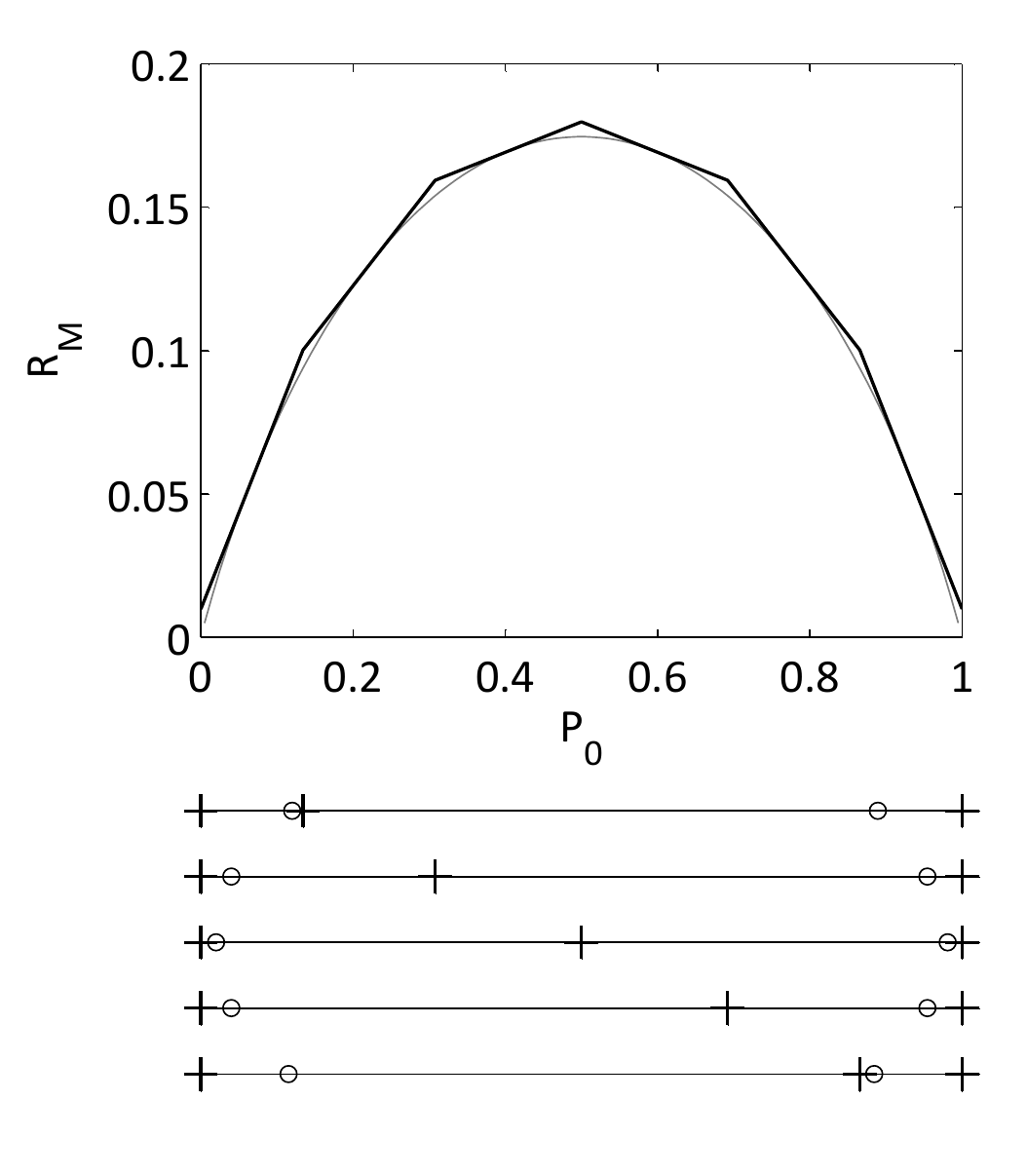}}
    \hfil
    \subfloat[$K=3$]{\includegraphics[width=1.8in]{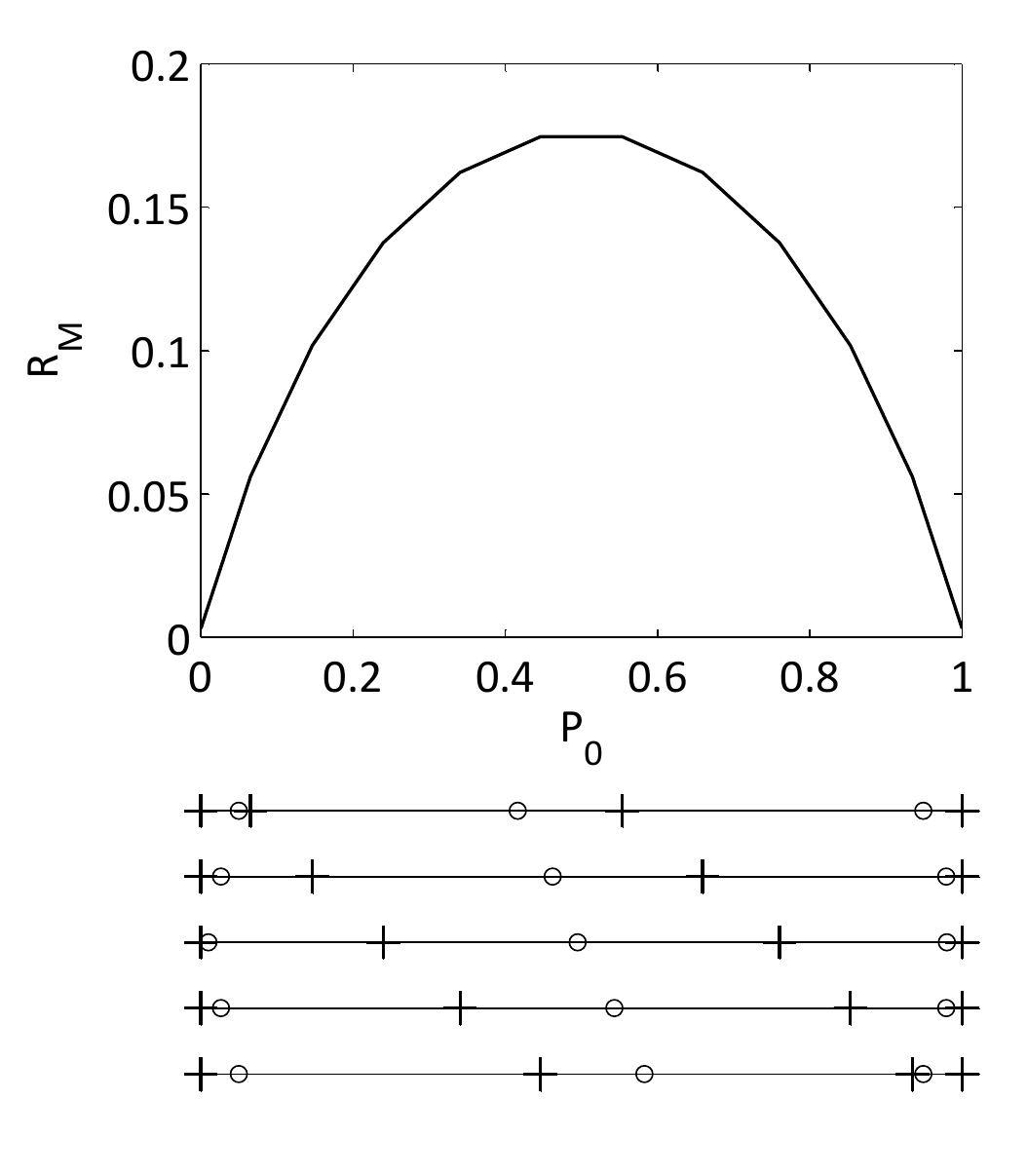}}
    \subfloat[$K=4$]{\includegraphics[width=1.8in]{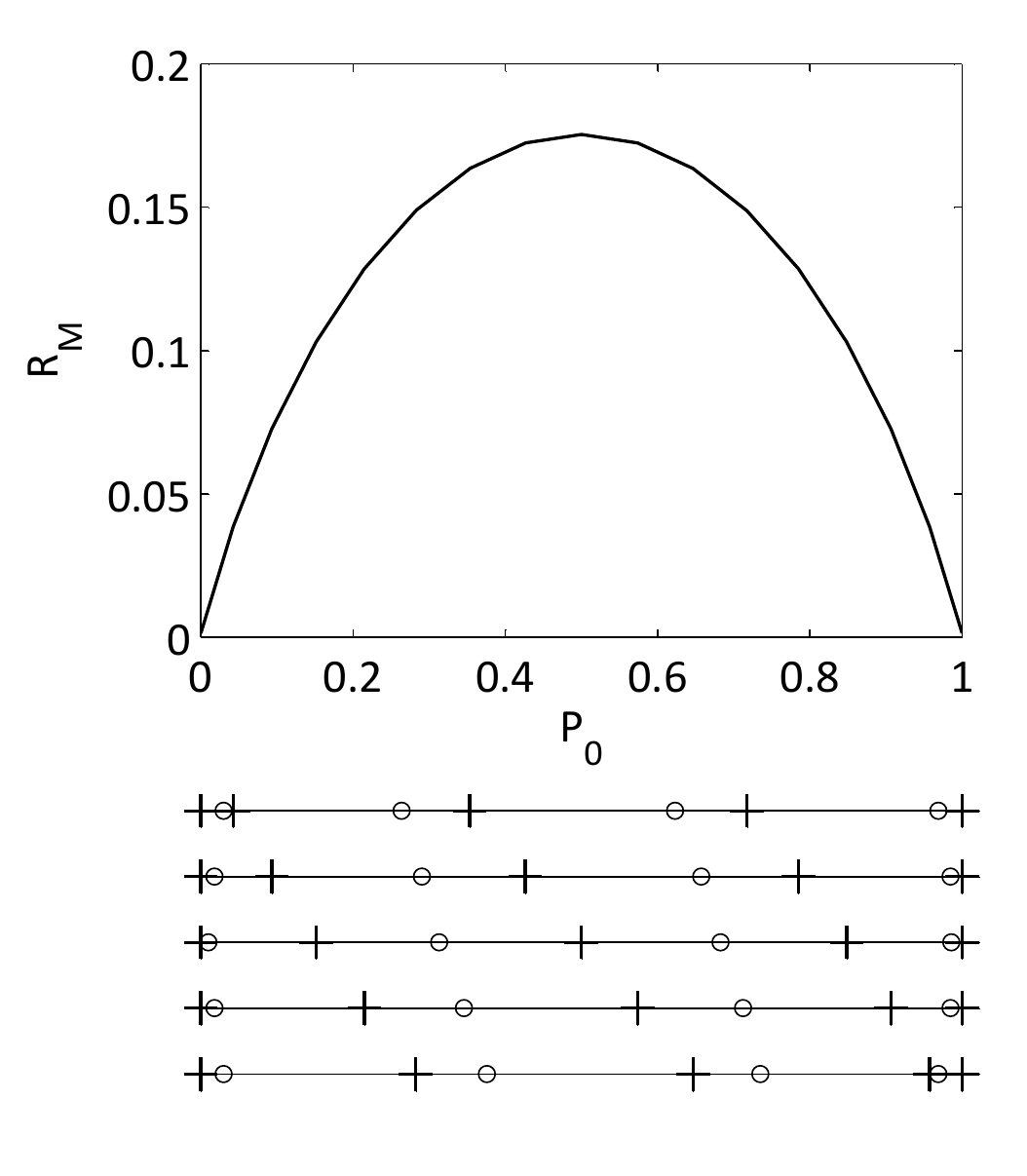}}
    }
    \caption{Optimal diverse $K$-level quantizers (cell boundaries as +'s and representation points as $\circ$'s) for uniformly distributed $P_0$ and the resulting mismatched Bayes risk when $N = 5$ agents perform distributed hypothesis testing fused by the {\sc majority} rule. The parameters are defined as $c_{10} = c_{01} = 1$, $\sigma = 1$, and $u_i = 1/5$, for $i = 1, \ldots, 5$. For comparison, the unquantized Bayes risk curve is depicted in gray in (a) and (b).}
    \label{fig:MajorityRuleQuantizer}
\end{figure}

\begin{figure}
  \centering{
    \subfloat[$K=1$]{\includegraphics[width=1.8in]{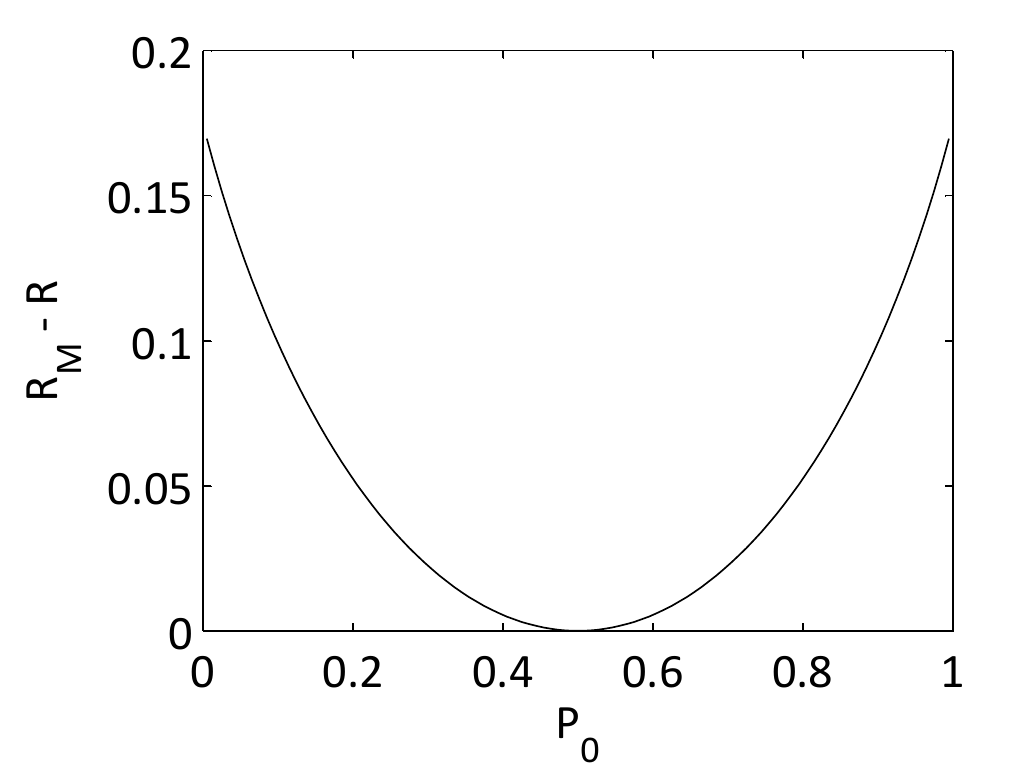}}
    \subfloat[$K=2$]{\includegraphics[width=1.8in]{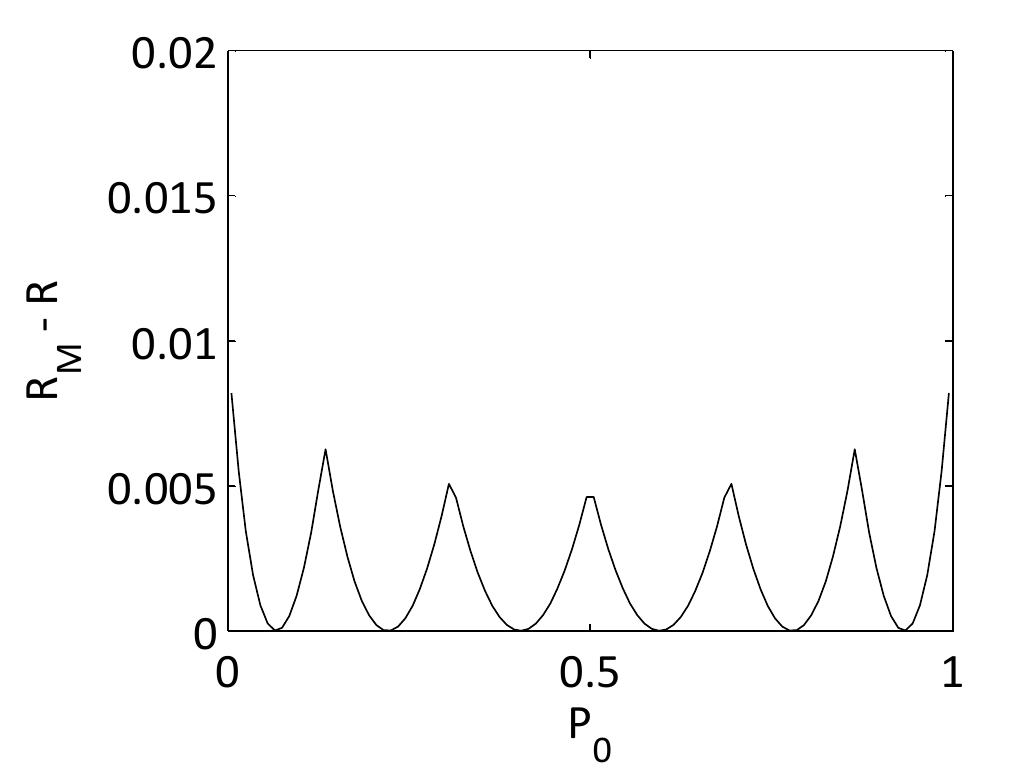}}
    \hfil
    \subfloat[$K=3$]{\includegraphics[width=1.8in]{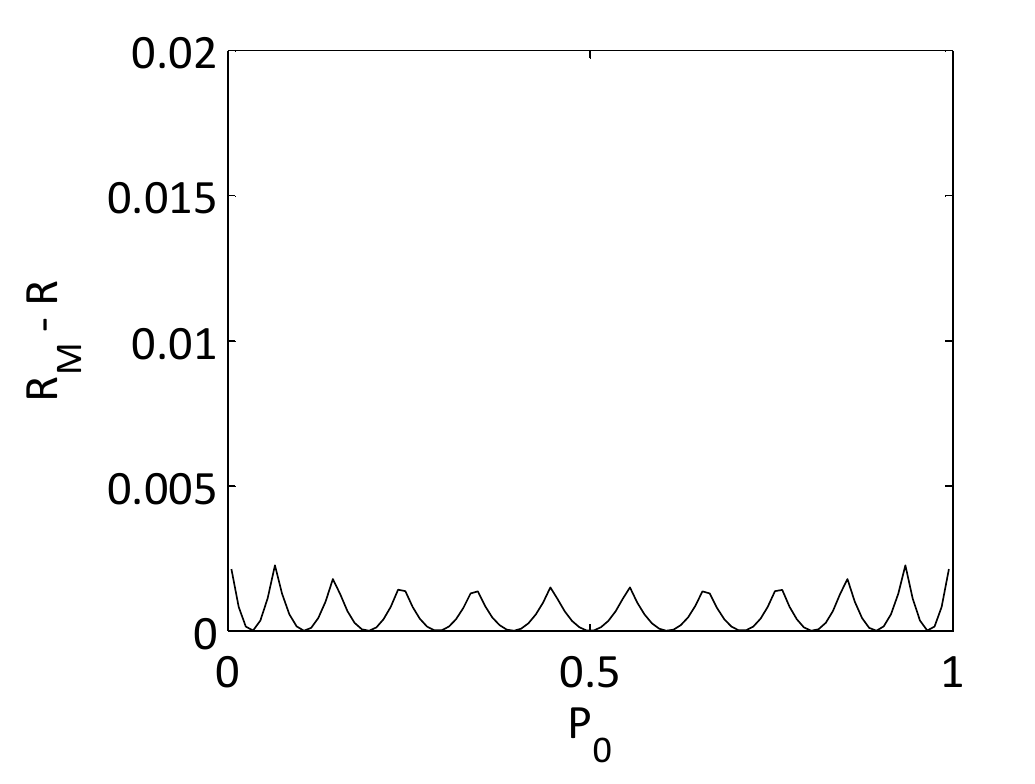}}
    \subfloat[$K=4$]{\includegraphics[width=1.8in]{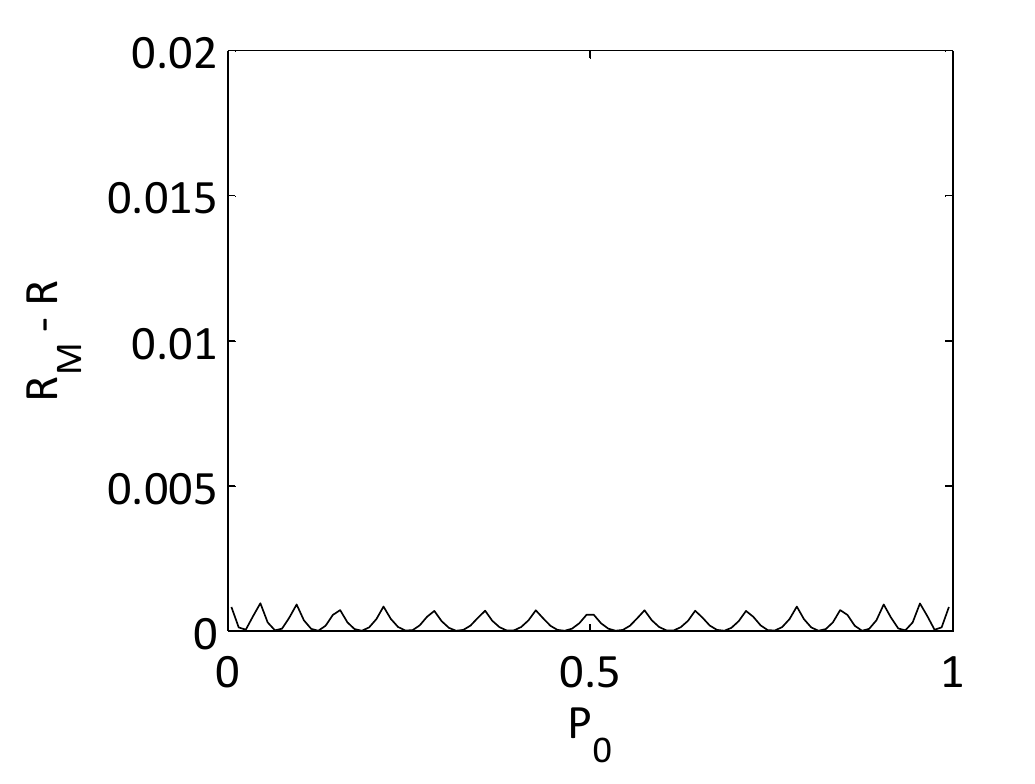}}
  }
  \caption{Bayes risk error of the quantizers in Fig.~\ref{fig:MajorityRuleQuantizer}.}
    \label{fig:MajorityRuleExcessBR}
\end{figure}

\begin{figure}
    \centering
    \includegraphics[width=3.2in]{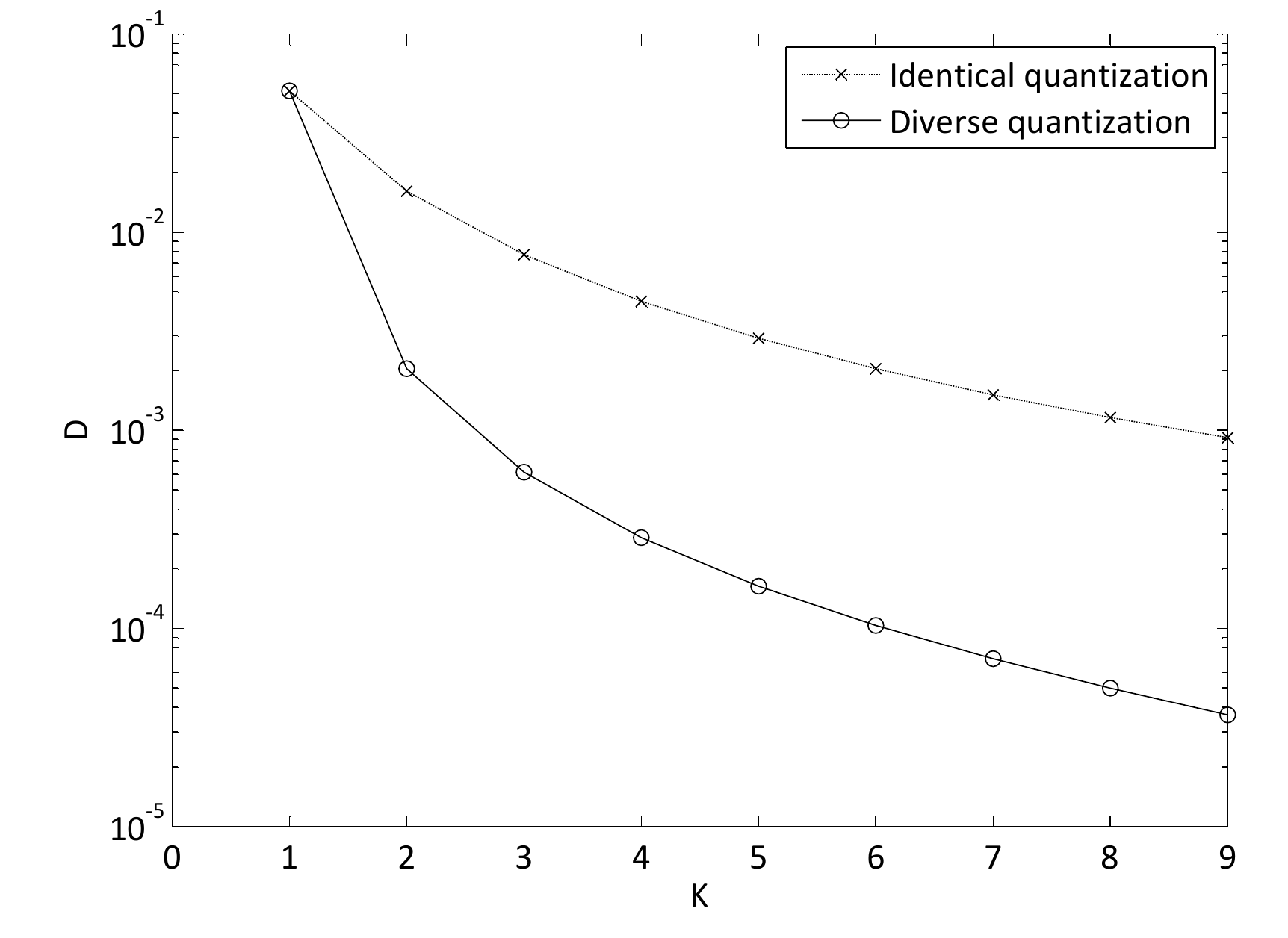}
    \caption{Mean Bayes risk error for uniformly distributed $P_0$ in the example of Fig.~\ref{fig:MajorityRuleQuantizer}.}
    \label{fig:MajorityRuleMBRE}
\end{figure}

Fig.~\ref{fig:MajorityRuleQuantizer} shows an example of a team of five agents\footnote{An example of a team of three agents is given in~\cite{RhimVG2011a}.}; it depicts an optimal set of diverse $K$-level quantizers and the resulting Bayes risk for $K = 1, \ldots, 4$. The Bayes risk error due to quantization of prior probabilities is depicted in Fig.~\ref{fig:MajorityRuleExcessBR}, which shows improvement of decision making as $K$ increases. Also, the mean Bayes risk error is given in Fig.~\ref{fig:MajorityRuleMBRE}; it shows the advantage of using optimal diverse quantizers against using optimal identical quantizers.

Asymptotically, $\zetaMajN$ decays inversely with $N$;
more precisely~\cite{VarshneyRVG2011},
\begin{equation}
\label{eq:zetaMajN}
\lim_{N \rightarrow \infty} N \cdot \zetaMajN = \frac{\pi}{2}.
\end{equation}
If the agents could share their observations rather than their local hard decisions, the team would base its decision on the sample mean of the observations.  The resulting performance is governed by the variance of the sample mean of the noise variables $W_1,\ldots,W_N$.  For $\sigma = 1$,
\[ \lim_{N \rightarrow \infty} N \cdot \var \left(N^{-1}{\textstyle\sum_{i = 1}^{N} W_i} \right) = 1.\]
The ratio of $\pi/2$ between the two asymptotic variances quantifies the loss in using majority vote rather than fusing full measurements.

\subsection{Gaussian Likelihoods, {\sc or} Rule}    \label{sec:OrRule}

Maintaining the Gaussian observation model from Section~\ref{sec:GaussianMajorityExample},
now consider fusion using the {\sc or} rule: the global decision is $h_0$ only when all agents declare $h_0$. The equivalent single-agent model has an additive noise $\VOrN$, which is the maximum order statistic of noises $W_1, \ldots, W_N$. The pdf of $\VOrN$ is given by
\[f_{\VorN}(v) = N F_{W}^{N - 1} (v) f_W(v).\]
The mean of $\VOrN$ is proportional to $\sigma$,
\[\E[\VOrN] = \muOrN \cdot \sigma,\]
and the variance of $\VOrN$ is proportional to $\sigma^2$,
\[\Var(\VOrN) = \zetaOrN \cdot \sigma^2;\]
these can be proven analogously to (\ref{eq:MajorityRuleVariance}).
The factor $\muOrN$ is increasing in $N$ but $\zetaOrN$ is decreasing in $N$. For example, $\muOr_3 = 0.8463$ and $\muOr_5 = 1.1630$; $\zetaOr_3 = 0.5595$ and $\zetaOr_5 = 0.4475$. For other $N$ between 1 and 20, the values of $\muOrN$ and $\zetaOrN$ are listed in~\cite{Teichroew1956}.

\begin{figure}
    \centering{
    \subfloat[$K=1$]{\includegraphics[width=1.8in]{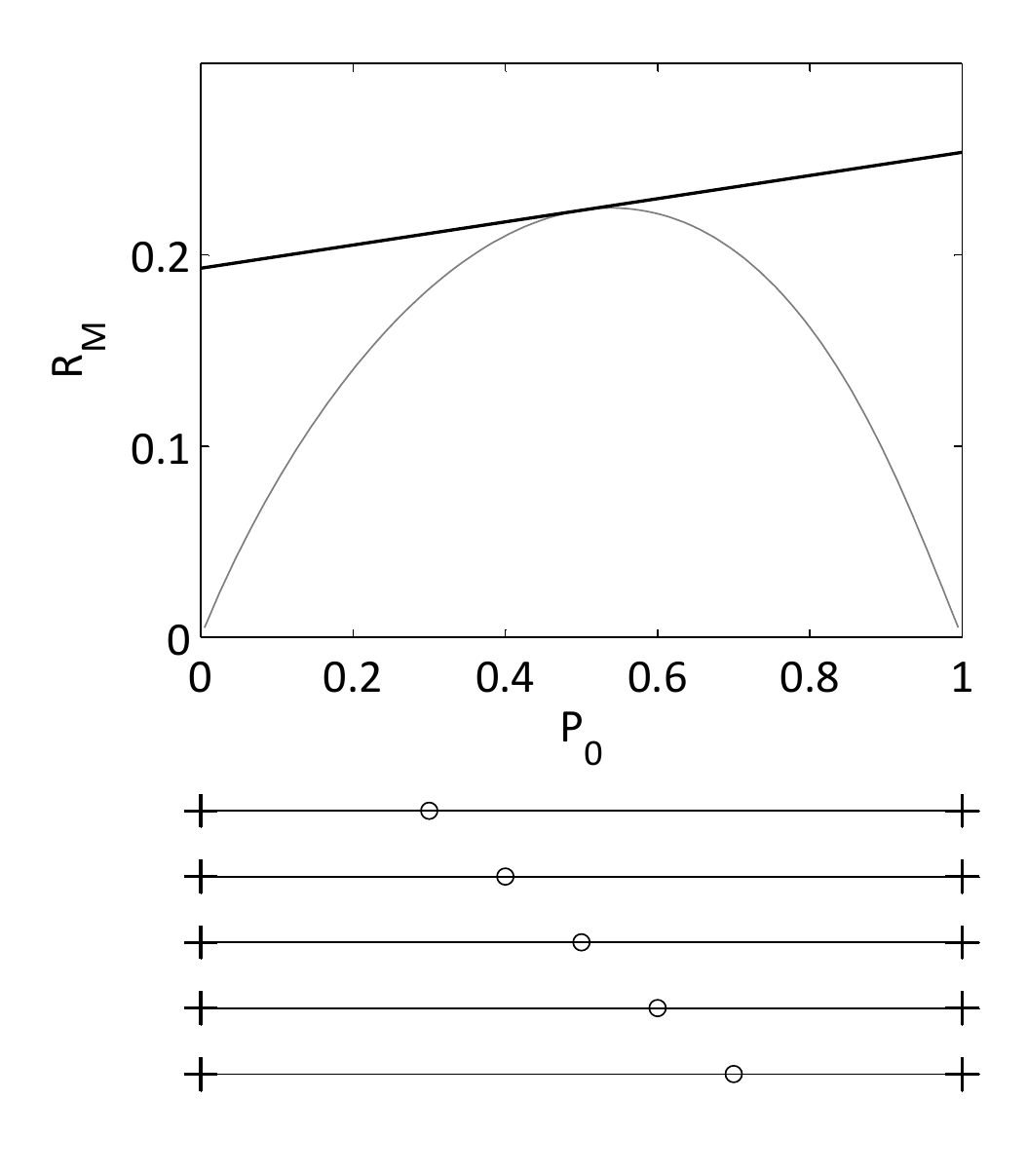}}
    \subfloat[$K=2$]{\includegraphics[width=1.8in]{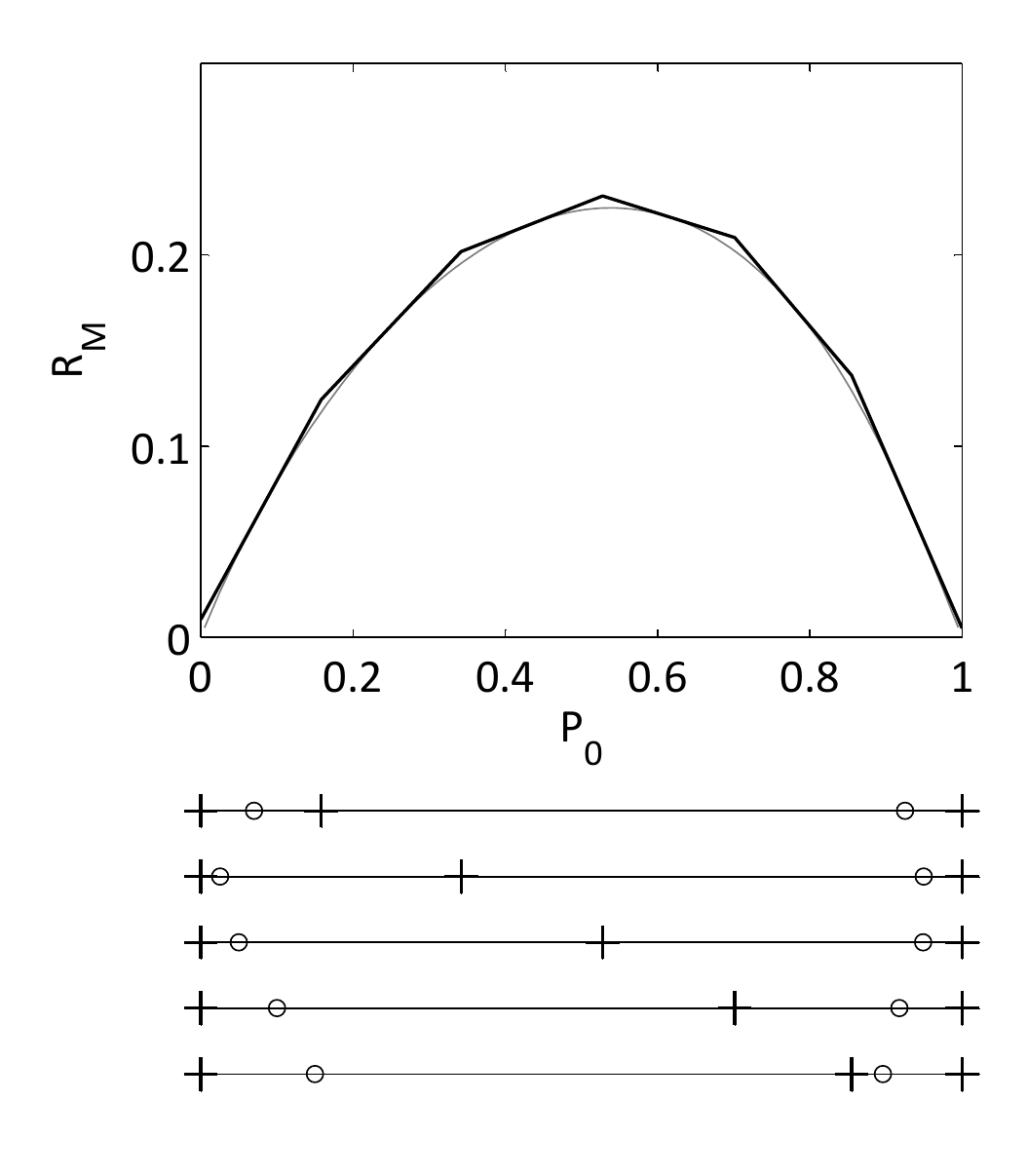}}
    \hfil
    \subfloat[$K=3$]{\includegraphics[width=1.8in]{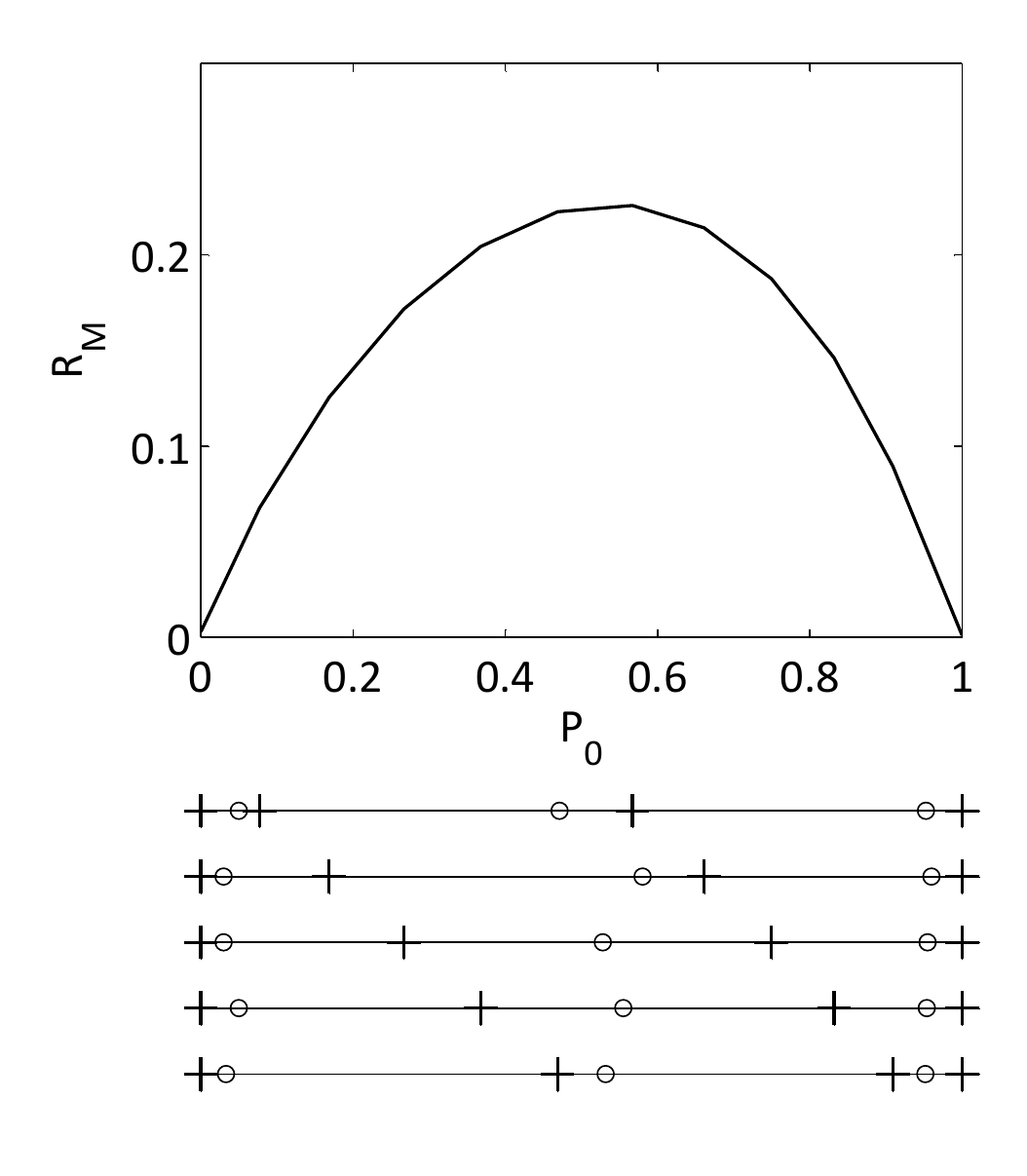}}
    \subfloat[$K=4$]{\includegraphics[width=1.8in]{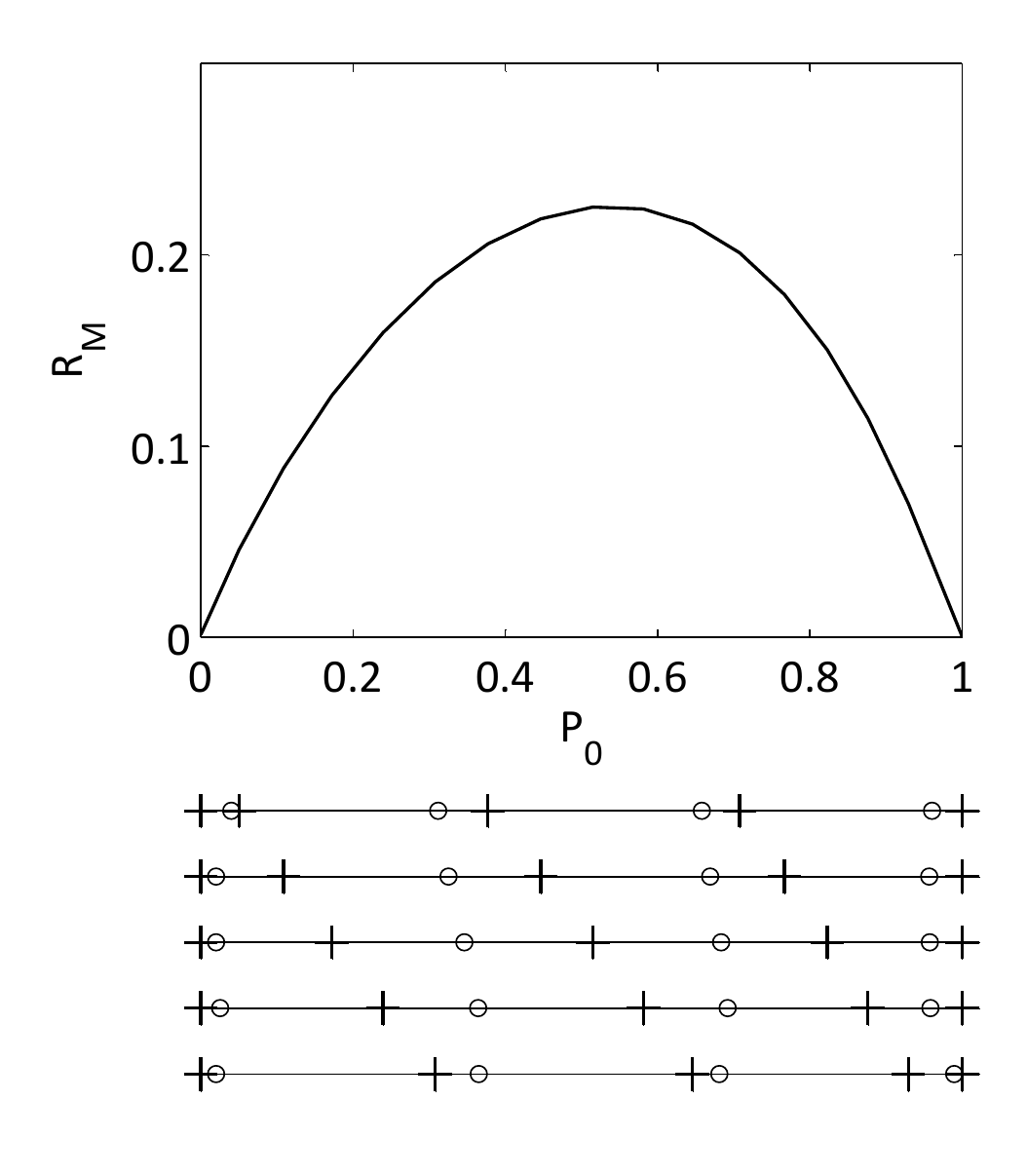}}
    }
    \caption{Optimal diverse $K$-level quantizers (cell boundaries as +'s and representation points as $\circ$'s) for uniformly distributed $P_0$ and the resulting mismatched Bayes risk when $N = 5$ agents perform distributed hypothesis testing fused by the {\sc or} rule. The parameters are defined as $c_{10} = c_{01} = 1$, $\sigma = 1$, and $u_i = 1/5$, for $i = 1, \ldots, 5$. For comparison, the unquantized Bayes risk curve is depicted in gray in (a) and (b).}
    \label{fig:OrRuleQuantizer}
\end{figure}

\begin{figure}
  \centering{
    \subfloat[$K=1$]{\includegraphics[width=1.8in]{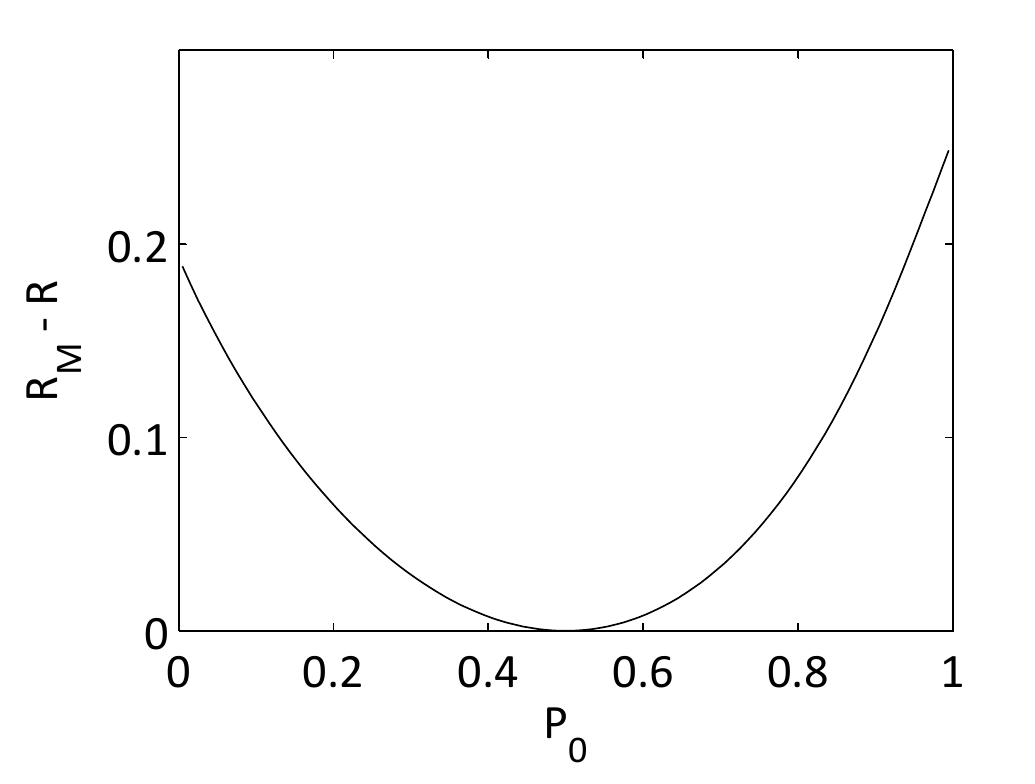}}
    \subfloat[$K=2$]{\includegraphics[width=1.8in]{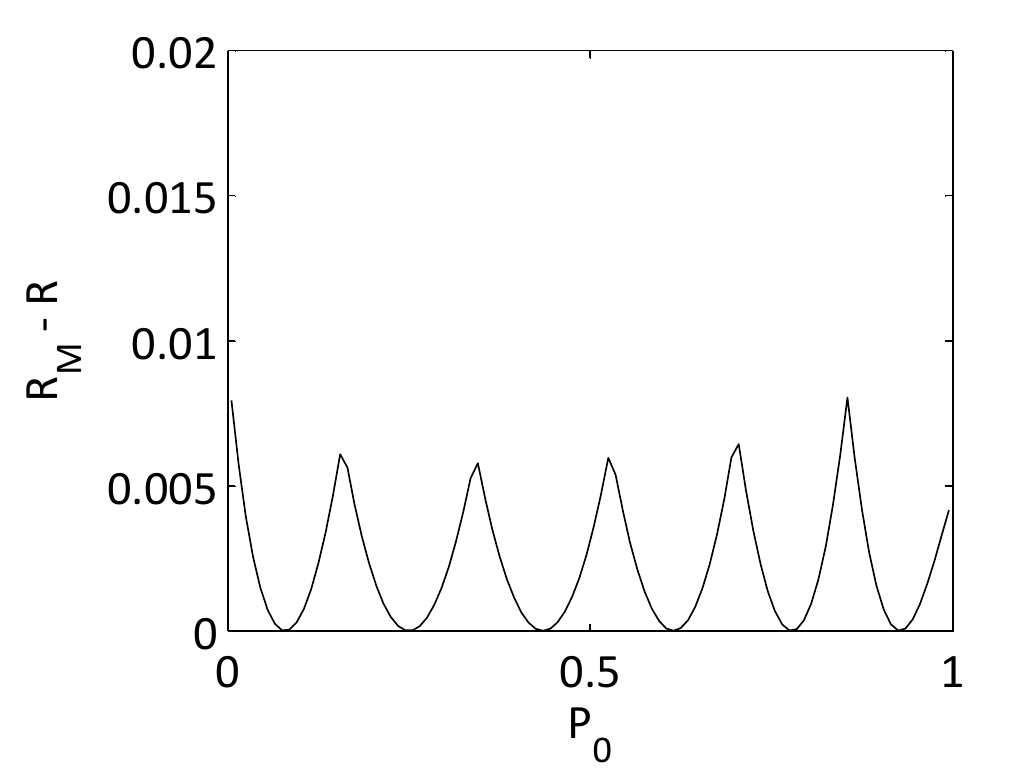}}
    \hfil
    \subfloat[$K=3$]{\includegraphics[width=1.8in]{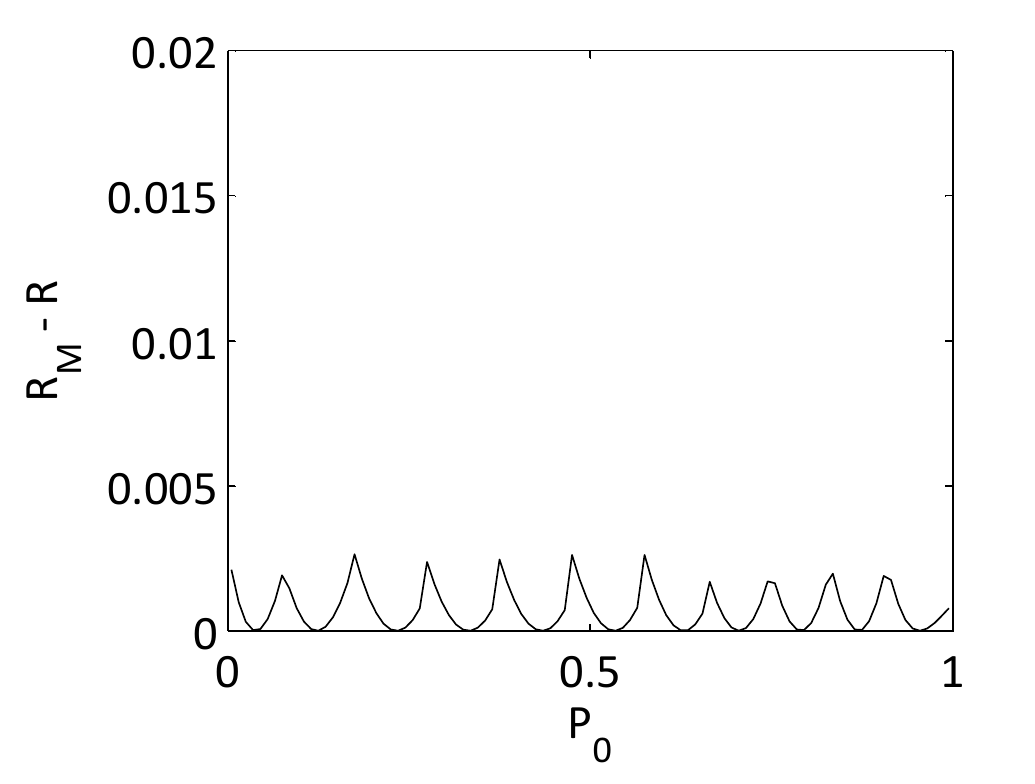}}
    \subfloat[$K=4$]{\includegraphics[width=1.8in]{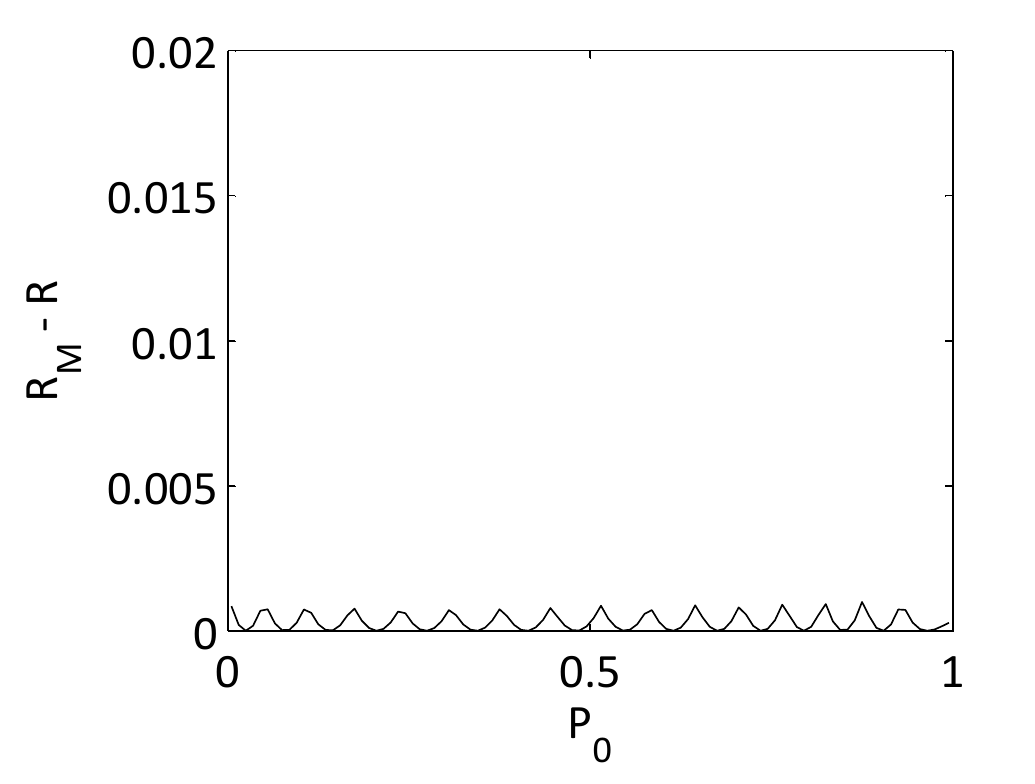}}
  }
  \caption{Bayes risk error of the quantizers in Fig.~\ref{fig:OrRuleQuantizer}.}
    \label{fig:OrRuleExcessBR}
\end{figure}

\begin{figure}
    \centering
    \includegraphics[width=3.2in]{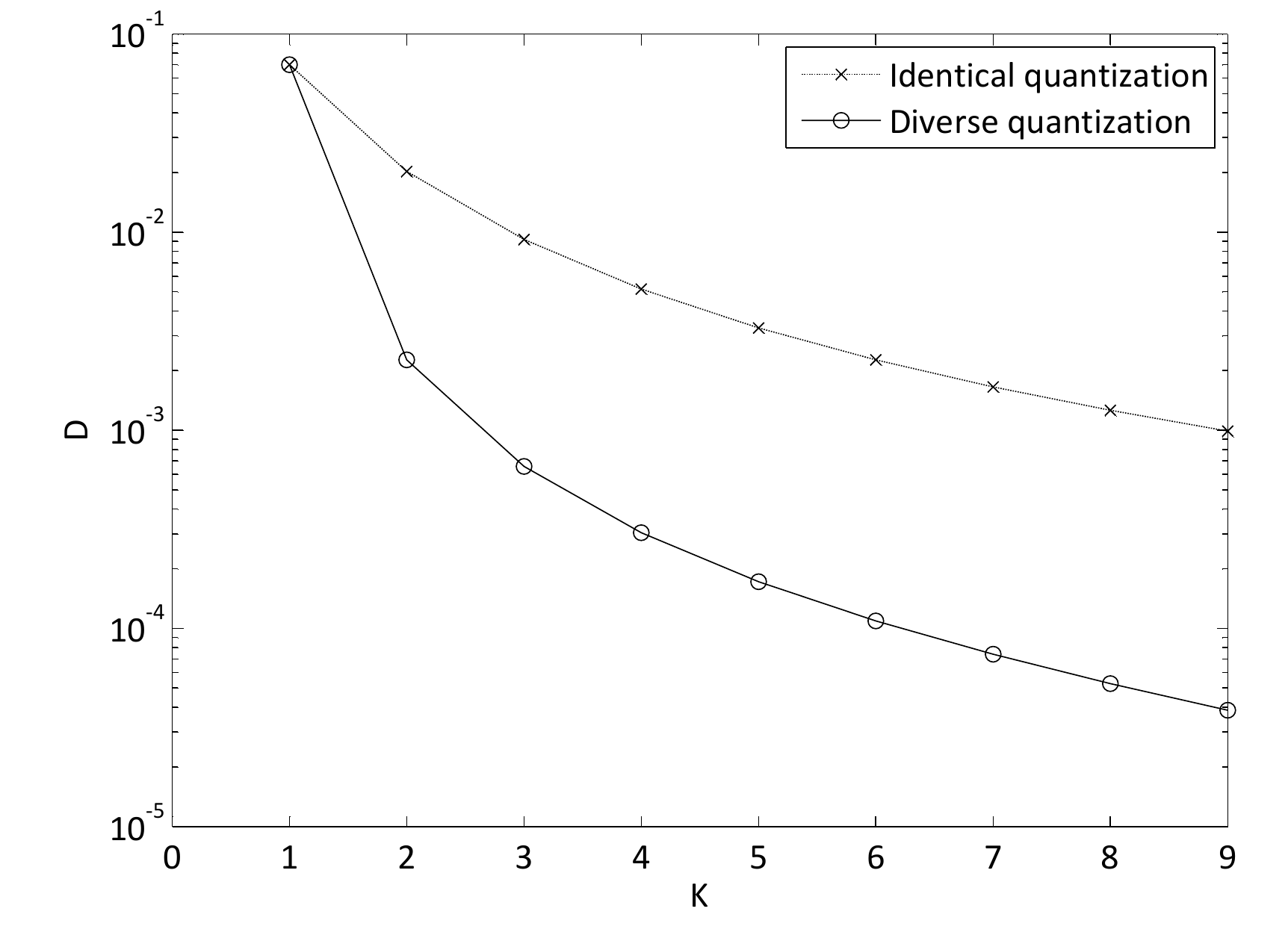}
    \caption{Mean Bayes risk error for uniformly distributed $P_0$ in the example of Fig.~\ref{fig:OrRuleQuantizer}.}
    \label{fig:OrRuleMBRE}
\end{figure}

An optimal set of diverse $K$-level quantizers for a team of five agents is given in Fig.~\ref{fig:OrRuleQuantizer} for $K = 1, \ldots, 4$. Fig.~\ref{fig:OrRuleExcessBR} and Fig.~\ref{fig:OrRuleMBRE} show the Bayes risk error and the mean Bayes risk error due to the quantization. The trends are similar to those in Section~\ref{sec:GaussianMajorityExample}, which considers the same model except for the fusion rule. The difference in the fusion rule changes the equivalent single-agent model. However, the same optimization algorithm of quantizers for prior probabilities can be applied in any case: designing the optimal identical $(N(K - 1) + 1)$-level quantizer of the equivalent single agent and disassembling the quantizer into $N$ diverse $K$-level quantizers. The algorithm comes from the relationship between the perceived Bayes risk and the common risk, which is defined as the weighted sum of the perceived Bayes risk. Thus, the algorithm does not depend on how the team of agents make decisions and the error probabilities $\GlobalI$ and $\GlobalII$ are computed.

Average performance with the {\sc or} rule (Fig.~\ref{fig:OrRuleMBRE}) is
slightly worse than with the {\sc majority} rule (Fig.~\ref{fig:MajorityRuleMBRE}).
A qualitative understanding can be obtained through study of $\VOrN$ and $\VMajN$,
and asymptotic behavior of the variances of these random variables suggests
that the performance gap increases with $N$.

\begin{figure}
  \centering{
    \subfloat[{\sc majority} rule]{\includegraphics[width=1.8in]{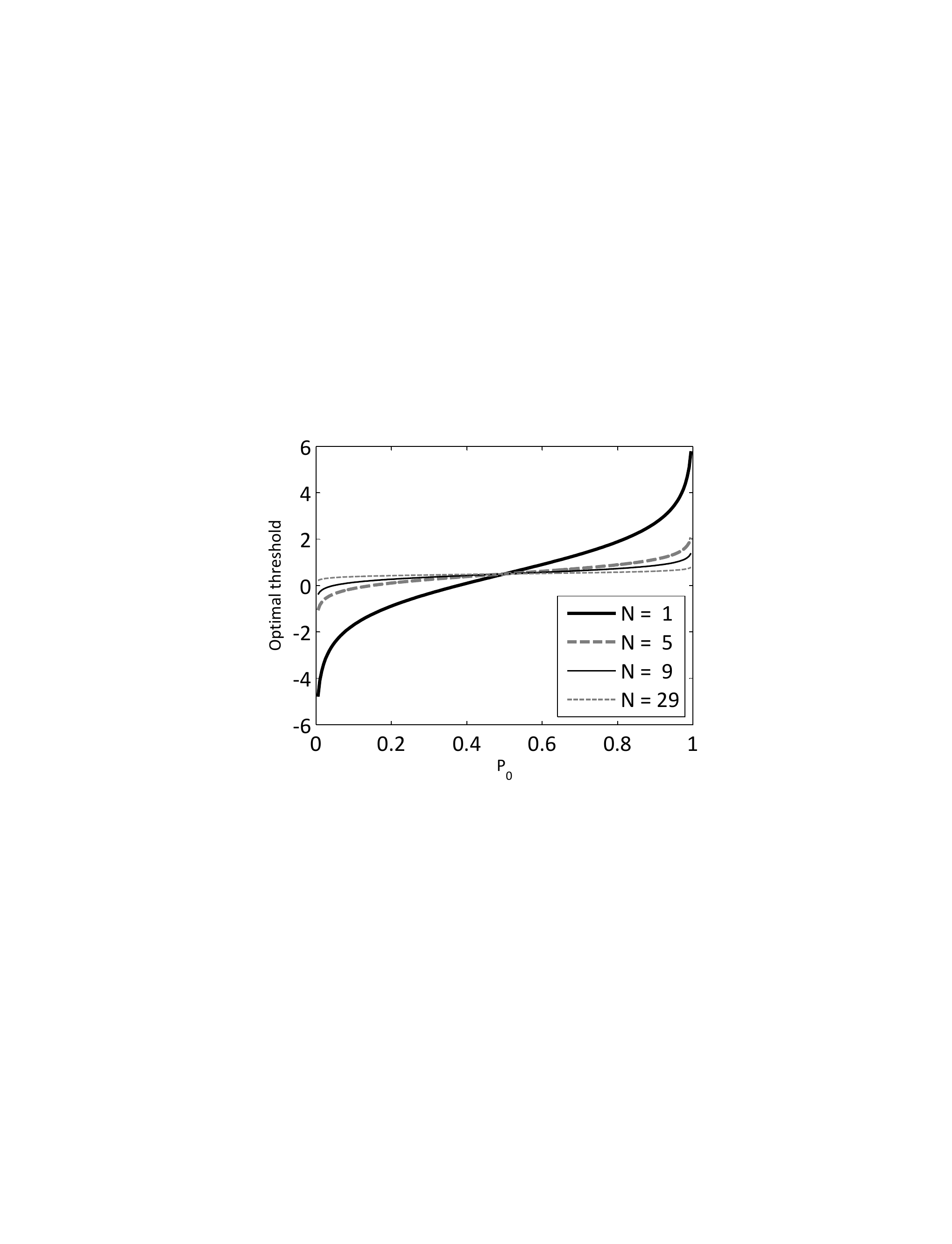}}
    \subfloat[{\sc or} rule]{\includegraphics[width=1.8in]{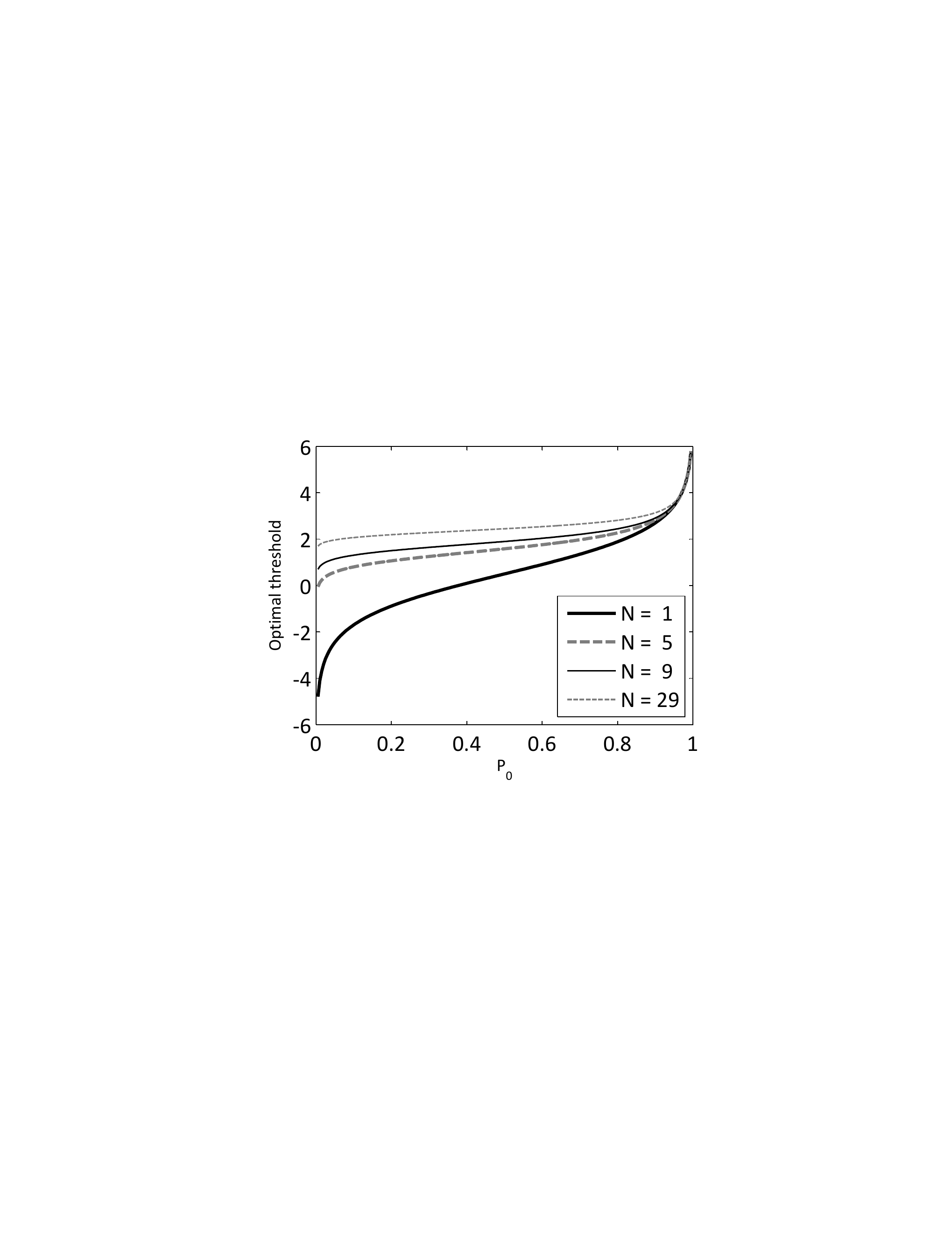}}
  }
  \caption{Optimal decision threshold for $N$ agents for $c_{10} = c_{01} = 1$ and $\sigma = 1$.}
    \label{fig:DecisionThreshold}
\end{figure}

While the mean of $\VMajN$ is zero for all $N$,
$\muOrN$ (the mean of $\VOrN$) is positive for $N \geq 2$ and increasing with $N$.
In the equivalent single-agent model under the {\sc or} rule,
the single agent tends to observe something larger than a true signal
by about $\muOrN$ because of the noise $\VOrN$.
Hence, the optimal decision threshold of the single agent is larger than
that in the model of Section~\ref{sec:GaussianMajorityExample}.
This can be interpreted as any individual agent optimally requiring
``stronger evidence'' to declare $h_1$ as the number of agents increases.
The optimal decision thresholds for $N=5$ are shown in Fig.~\ref{fig:DecisionThreshold}.
The asymptotic growth of $\muOrN$ is given by
(see~\cite[Ex.~10.5.3]{DavidNagaraja2003})
\[
  \lim_{N \rightarrow \infty} \muOrN / (2\log N)^{1/2} = 1.
\]

If all the equivalent noise variables had the same standardized distribution,
then {\sc majority} and {\sc or} fusion rules for fixed $N$
could be compared through the variances $\zetaMajN$ and $\zetaOrN$.
We saw in \eqref{eq:zetaMajN} that $\zetaMajN = \Theta(1/N)$.
The decay of $\zetaOrN$ is much slower;
specifically (see~\cite[Ex.~10.5.3]{DavidNagaraja2003}),
\[
  \lim_{N \rightarrow \infty} \left({\textstyle\frac{12}{\pi^2}} \log N\right) \cdot \zetaOrN = 1.
\]
This suggests that for large $N$ (and the Gaussian likelihood case), the {\sc majority} rule is more effective than the {\sc or} rule.

\subsection{Exponential Likelihoods}    \label{sec:ExponentialExample}
Consider a particle that disappears with rate $s_0$ in state $h_0$ and with rate $s_1$ in state $h_1$, where $s_0 > s_1$.
Conditioned on $H$,
the particle has an exponentially-distributed lifetime $Y$:
\begin{equation*}
    f_{Y | H}(y \MID h_m) = s_m e^{-s_m y}.
\end{equation*}
An agent observes that the particle disappears at time $Y = y$ and performs Bayesian hypothesis testing by the likelihood ratio test:
\begin{equation*}
    \frac{f_{Y | H}(y \MID h_1)}{f_{Y | H}(y \MID h_0)} \overset{\Hhat = h_1}{\underset{\Hhat = h_0}{\gtreqless}} \frac{p_0 c_{10}}{(1 - p_0) c_{01}}.
\end{equation*}
The likelihood ratio test can be simplified to the decision rule
\begin{equation*}
    y \overset{\Hhat = h_1}{\underset{\Hhat = h_0}{\gtreqless}} \frac{1}{s_0 - s_1} \log \left( \frac{s_0}{s_1} \frac{p_0 c_{10}}{(1 - p_0) c_{01}} \right) = \lambda,
\end{equation*}
which yields errors with probabilities
\begin{equation*}
    P_{e}^{\rm I}  = e^{-s_0 \lambda}
\qquad
\mbox{and}
\qquad
    P_{e}^{\rm II}  = 1 - e^{-s_1 \lambda}.
\end{equation*}

Now suppose that $N$ agents perform Bayesian hypothesis testing with $N$ particles in the same state; Agent $i$ observes that the $i$th particle disappears at time $Y_i = y_i$ and applies a common decision threshold $\lambda$ to its observation to make its local decision. All agents' decisions are fused by $L$-out-of-$N$ rule. Then the equivalent single-agent model is to consider a particle with lifetime $Y_{(L)}$, which is the $L$th longest lifetime among $\{Y_1, \ldots, Y_N\}$,
\begin{align}
    & f_{Y_{(L)} | H}(y \MID h_m) \nonumber \\ \nonumber
    & = \frac{N!}{(N - L)! (L - 1)!} (1 - e^{-s_m y})^{N - L} (e^{-s_m y})^{L - 1} s_m e^{-s_m y}.
\end{align}
The probabilities of global errors are
\begin{align}
    \GlobalI & = \sum_{n = M}^{N} {N \choose n} (e^{-s_0 \lambda})^n (1 - e^{-s_0 \lambda})^{N - n}, \nonumber \\
    \GlobalII & = \sum_{n = N - M + 1}^{N} {N \choose n} (1 - e^{-s_1 \lambda n})^{n} (e^{-s_1 \lambda})^{N - n}. \nonumber
\end{align}

\begin{figure}
  \centering{
    \subfloat[]{\includegraphics[width=1.8in]{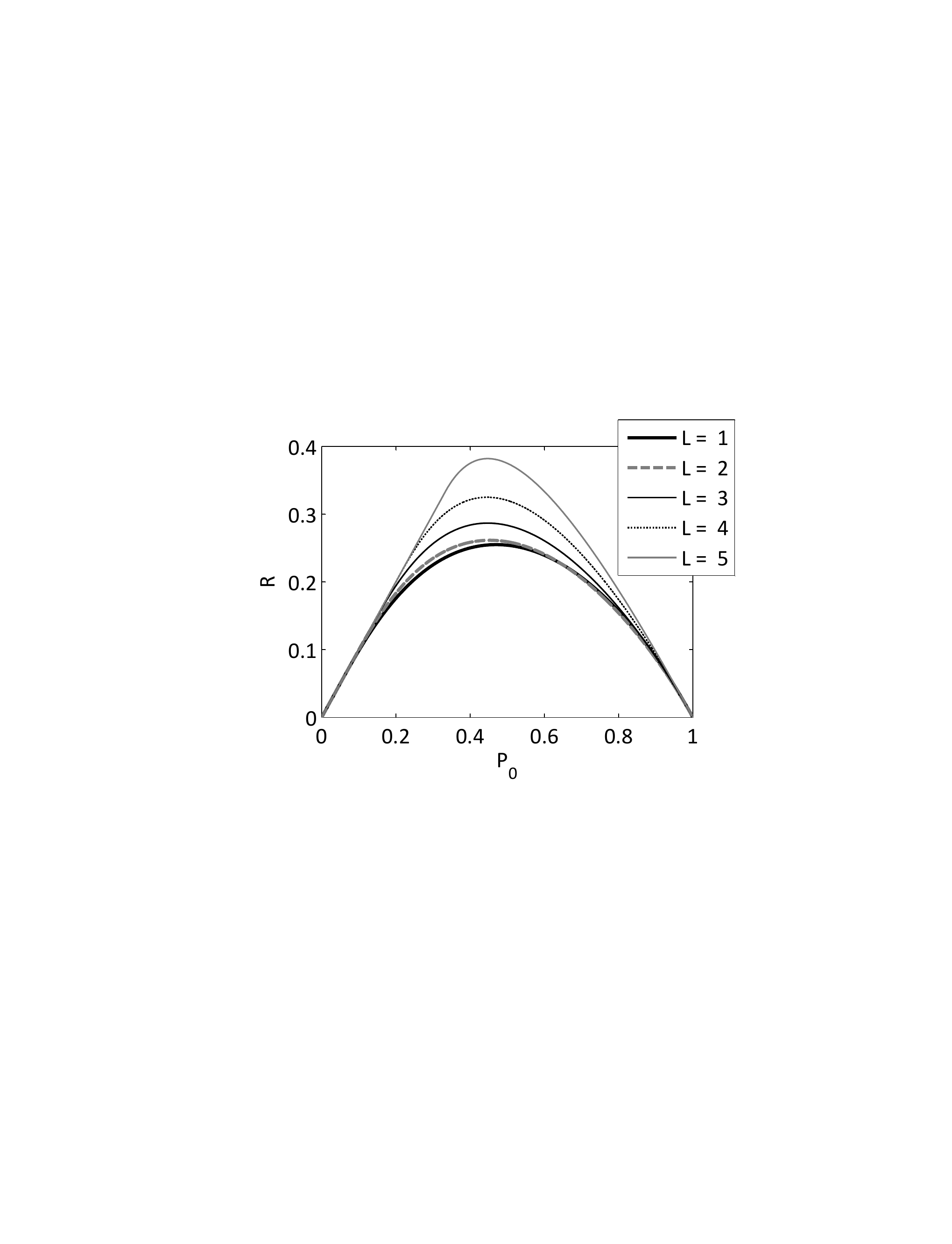}%
    \label{fig:BR_Exponential}}
    \subfloat[]{\includegraphics[width=1.8in]{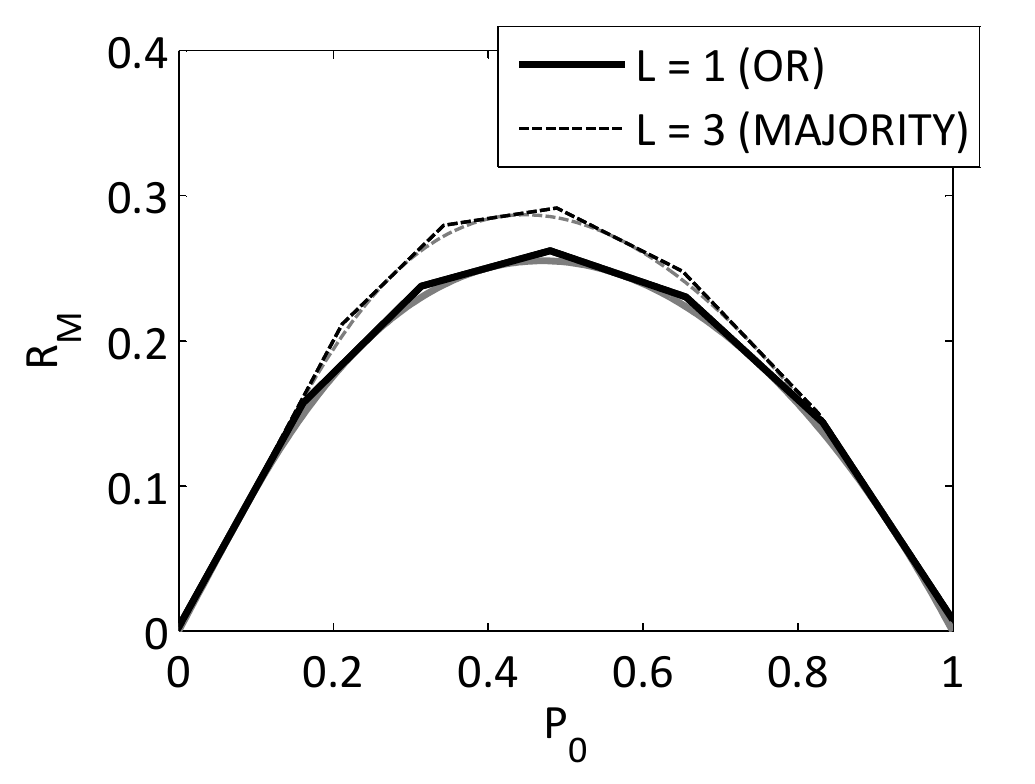}%
    \label{fig:MismatchedBR_Exponential}}
  }
  \caption{Bayes risk (a) without quantization of prior probabilities and (b) with diverse 2-level minimum MBRE quantizers for $N = 5$, $s_0 = 2$, $s_1 = 1$, $c_{10} = c_{01} = 1$, and $u_i = 1 /5 $ for $i = 1, \ldots, 5$.}
    \label{fig:Exponential}
\end{figure}

In this scenario, a small $L$ is a good choice for the fusion rule.  Fig.~\ref{fig:BR_Exponential} shows that the smallest mean Bayes risk without quantization of prior probabilities is achieved by $L = 1$.  Fig.~\ref{fig:MismatchedBR_Exponential} depicts an example of Bayes risk when five agents use diverse minimum MBRE quantizers and their fusion rule is the {\sc or} rule or the {\sc majority} rule.

\section{Minimax Bayes Risk Error Quantizers}    \label{sec:Minimax}

Rather than mean Bayes risk error, let us consider maximum Bayes risk error as the criterion for optimizing quantizer design~\cite{VarshneyVarshney2011}.  Such quantizers maximize worst case performance, whereas minimum MBRE quantizers maximize average performance. The minimax Bayes risk error quantizer is defined by the following optimization problem:
\begin{equation*}
    (q^{*}_1, \ldots, q^{*}_N) = \argmin_{(q_1, \ldots, q_N)} \max_{p_0} d(p_0, q_1(p_0), \ldots, q_N(p_0)).
\end{equation*}

The minimax Bayes risk error quantizer has the same nearest neighbor condition as the minimum MBRE quantizer for $N = 1$.
On the other hand, a centroid condition for optimality of a regular quantizer for $N = 1$ is different from that of the minimum MBRE quantizer~\cite{VarshneyVarshney2011}.  For any quantization point $a$, the Bayes risk error $d(p_0, a) = R_M(a) - R(p_0)$ is nonnegative and strictly convex in $p_0$, with minimum value of zero attained only at $p_0 = a$.  Thus, its maximum point within its $k$th cell $\mathcal{R}_k = [b_{k - 1}, b_k)$ is a cell boundary: $b_{k - 1}$ or $b_k$.  The point $a^{*}$ that satisfies
\begin{equation}
    d(b_{k - 1}, a^{*}) = d(b_k, a^{*})
    \label{eq:MinimaxCentroid}
\end{equation}
minimizes the maximum Bayes risk error within $\mathcal{R}_k$.  This is the centroid condition for the representation point $a_k = a^{*}$ of cell $\mathcal{R}_k$. The minimax Bayes risk error quantizer can be found by alternatively applying the nearest neighbor condition (\ref{eq:NearestNeighbor}) and the centroid condition (\ref{eq:MinimaxCentroid}) through the iterative Lloyd--Max algorithm.

For $N > 1$, agents can take advantage of diversity in the same way as in Section~\ref{sec:DiverseQuantization}.  A team of $N$ agents bonded by the $L$-out-of-$N$ fusion rule is equivalent to a single agent with noise equal to their $L$th largest noise $W_{(L)}$ (cf. Theorem~\ref{thm:EquivalentModel}).  Thus, identical quantizers that minimize maximum Bayes risk error of the $N$ agents also minimize that of the single agent and vice versa (cf. Theorem~\ref{thm:IdenticalQuantizer}).

Furthermore, if the $N$ agents collaborate by sharing the perceived common risk, there exists a set of $N$ diverse $K$-level quantizer that leads to the same Bayes risk for any $p_0$ as a set of identical $(N (K - 1) + 1)$-level quantizers does (cf. Theorem~\ref{thm:EquivalentQuantizer}).  Therefore, the identical $(N (K - 1) + 1)$-level minimax Bayes risk error quantizers can be transformed into $N$ diverse $K$-level minimax Bayes risk error quantizers (cf. Theorem~\ref{thm:GuaranteedLevel}).

\begin{figure}
  \centering{
    \subfloat[Bayes risk]{\includegraphics[width=1.8in]{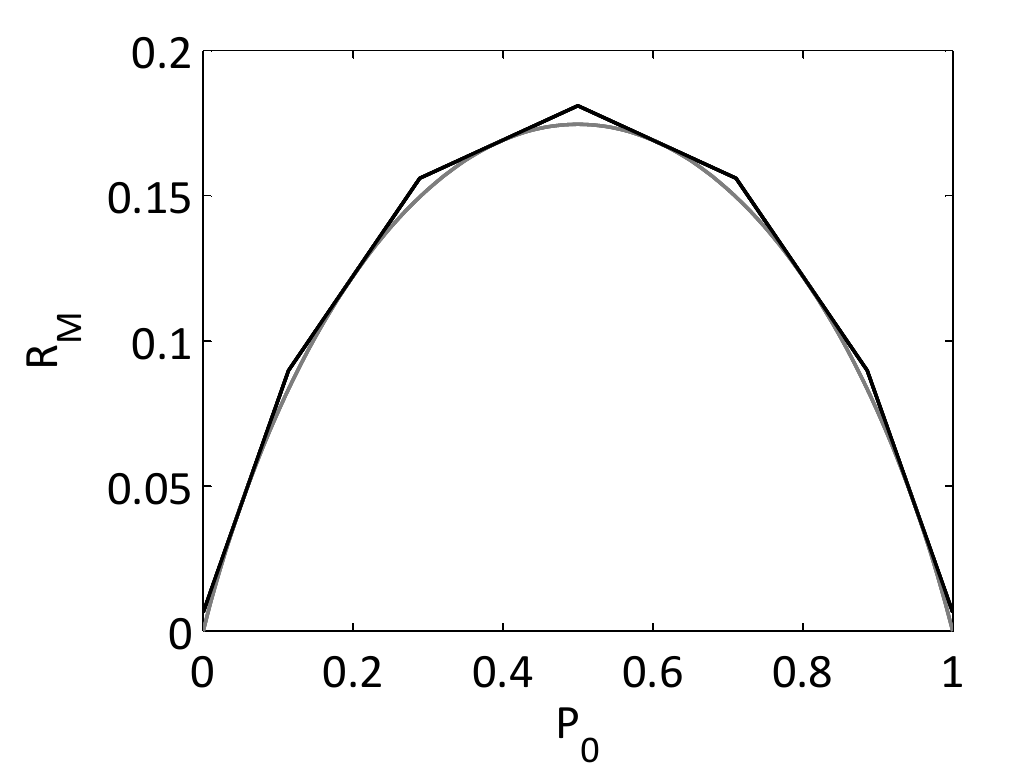}%
    \label{fig:MinimaxBR}}
    \subfloat[Bayes risk error]{\includegraphics[width=1.8in]{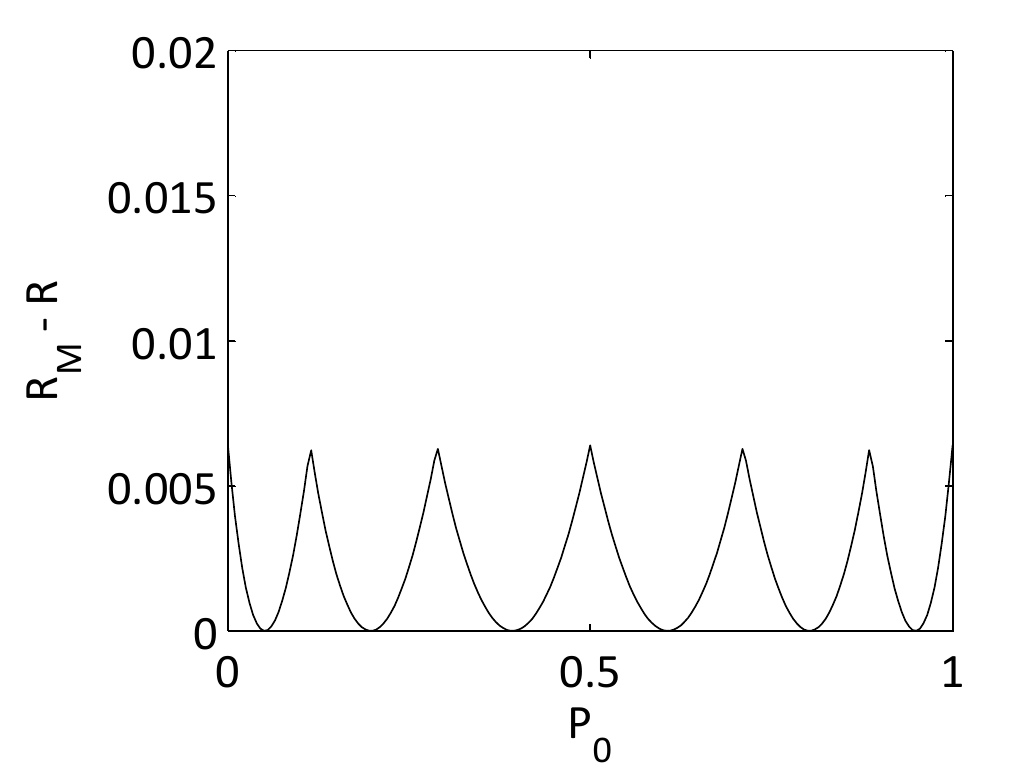}%
    \label{fig:MinimaxBRE}}
  }
  \caption{(a) Bayes risk and (b) Bayes risk error for $N = 5$ agents performing distributed hypothesis testing fused by the {\sc majority} rule. They observe signals corrupted by iid additive Gaussian noise $\mathcal{N}(0, \sigma^2)$. The parameters are defined as $s_0 = 0$, $s_1 = 1$, $c_{10} = c_{01} = 1$, $\sigma = 1$, and $u_i = 1/5$ for $i = 1, \ldots, 5$.}
    \label{fig:Minimax}
\end{figure}

Fig.~\ref{fig:Minimax} shows an example of Bayes risk when five agents use diverse 2-level minimax Bayes risk error quantizers and their fusion rule is {\sc majority}.  A property of minimax Bayes risk error quantizers is that the maximum Bayes risk errors at all cell boundaries are the same, which is shown in Fig.~\ref{fig:MinimaxBRE}.  In addition, the minimax Bayes risk error quantizers are not dependent on the distribution of $P_0$ as long as $f_{P_0}(p_0) > 0$ for all $p_0 \in [0, 1]$ because they minimize the worst case error, not the average.

\section{Conclusion}
\label{sec:Conclusion}

We have discussed distributed detection and data fusion performed by a team of agents when there is a distribution of prior probabilities and the agents only know quantized versions of prior probabilities. We have focused on how to take advantage of diversity in quantization. When all agents use identical quantizers, they are affected by the same perceived Bayes risk and the distributed hypothesis testing problem can be analyzed by existing theorems of decision theory. On the contrary, when they do not use identical quantizers, then they consider different perceived Bayes risks, which prevents them from collaborating in hypothesis testing. We let the agents use the perceived common risk as a new distortion measure of hypothesis testing so as to unite them as a team to perform distributed hypothesis testing in any case.

We have defined mean Bayes risk error as the optimization criterion for prior-probability quantizers. We have presented theorems to show that diverse quantizers are better than identical quantizers. The equivalence between multiple-agent decision making and single-agent decision making simplifies a team of agents that use identical quantizers to a single agent. By combining the equivalence theorem with the equivalence between diverse quantizers and identical quantizers used by a team of agents, we can take advantage of the study of the minimum MBRE quantizer of a single agent to analyze optimal diverse quantization for multiple agents.
It is shown that if the agents use diverse $K$-level quantizers, it has the same effect of using identical $(N(K - 1) + 1)$-level quantizers.

The equivalence theorems hold under the condition that all agents collaborate to perform Bayesian hypothesis testing. Hence, for any given distortion function of quantizers, diverse quantizers that minimize the distortion can be easily found.
While the case of minimum MBRE was covered in detail, we also discussed the
minimax Bayes risk error criterion.  In the minimax case, one can again
first design an $(N(K-1)+1)$-level quantizer for a single agent and then
disassemble the quantizer into $N$ diverse $K$-level quantizers.

\bibliographystyle{IEEEtran}
\bibliography{rhim_lib}
\end{document}